\documentclass[fleqn,usenatbib]{mnras}

\usepackage{newtxtext,newtxmath}

\usepackage{amsmath,bm}

\usepackage[T1]{fontenc}

\DeclareRobustCommand{\VAN}[3]{#2}
\let\VANthebibliography\thebibliography
\def\thebibliography{\DeclareRobustCommand{\VAN}[3]{##3}\VANthebibliography}

\usepackage{graphicx}	
\usepackage{bm}
\usepackage{amsmath}	

\usepackage{amssymb}	
\usepackage{multicol, blindtext} 

\usepackage{booktabs}

\defcitealias{trindadefalcao2020a}{Paper I}

\title[\textit{HST} Observations of {[O~III]} Emission in Nearby QSO2s]{Hubble Space Telescope [O~III] Emission-Line Kinematics in Two Nearby QSO2s: A Case for X-ray Feedback}

\author[Trindade Falcão et al.]{
Anna Trindade Falcão ,$^{1}$\thanks{E-mail: anna.trindade04@gmail.com}
S. B. Kraemer,$^{1}$
T. C. Fischer,$^{2}$
D. M. Crenshaw,$^{3}$
M. Revalski,$^{4}$
\newauthor
H. R. Schmitt,$^{5}$
W. P. Maksym,$^{6}$
M. Vestergaard,$^{7,8}$
M. Elvis,$^{9}$
C. M. Gaskell,$^{10}$
F. Hamann,$^{11}$
\newauthor
L. C. Ho,$^{12}$
J. Hutchings,$^{13}$
R. Mushotzky,$^{14}$
H. Netzer,$^{15}$
T. Storchi-Bergmann,$^{16}$
T. J. Turner,$^{17}$
\newauthor
M. J. Ward$^{18}$
\\
$^{1}$Institute for Astrophysics and Computational Sciences, Department of Physics, The Catholic University of America, Washington, DC 20064, USA\\
$^{2}$AURA for ESA, Space Telescope Science Institute, 3700 San Martin Drive, Baltimore, MD 21218, USA\\
$^{3}$Department of Physics and Astronomy, Georgia State University, Astronomy Offices, 25 Park Place, Suite 600, Atlanta, GA 30303, USA\\
$^{4}$Space Telescope Science Institute, 3700 San Martin Drive, Baltimore, MD 21218, USA\\
$^{5}$Naval Research Laboratory, Washington, DC 20375, USA\\
$^{6}$Harvard-Smithsonian Center for Astrophysics, 60 Garden St., Cambridge, MA 02138, USA\\
$^{7}$DARK, Niels Bohr Institute, University of Copenhagen, Jagtvej 128, 2200 Copenhagen N, Denmark\\
$^{8}$Steward Observatory and Department of Astronomy, University of Arizona, 933 N. Cherry Avenue, Tucson AZ 85721\\
$^{9}$Harvard-Smithsonian Center for Astrophysics, 60 Garden St., Cambridge, MA 02138, USA\\
$^{10}$Department of Astronomy and Astrophysics, University of California, Santa Cruz, CA 95064, USA\\
$^{11}$Department of Physics and Astronomy, University of California, Riverside, CA 92507, USA\\
$^{12}$Kavli Institute for Astronomy and Astrophysics, Peking University; School of Physics, Department of Astronomy, Peking University; Beijing 100871, China\\
$^{13}$Dominion Astrophysical Observatory, NRC Herzberg Institute of Astrophysics, 5071 West Saanich Road, Victoria, BC, V9E 2E7, Canada\\
$^{14}$Department of Astronomy, University of Maryland, College Park, MD 20742, USA\\
$^{15}$School of Physics and Astronomy, Tel Aviv University, Tel Aviv 69978, Israel\\
$^{16}$Departamento de Astronomia, Universidade Federal do Rio Grande do Sul, IF, CP 15051, 91501-970 Porto Alegre, RS, Brazil\\
$^{17}$Eureka Scientific, Inc., 2452 Delmer Street, Suite 100, Oakland, CA 94602-3017, USA\\
$^{18}$Centre for Extragalactic Astronomy, Department of Physics, University of Durham, South Road, Durham DH1 3LE, UK\\
}

\date{Accepted XXX. Received YYY; in original form ZZZ}

\pubyear{2021}

\begin{document}
\label{firstpage}
\pagerange{\pageref{firstpage}--\pageref{lastpage}}
\maketitle

\begin{abstract}
We present a dynamical study of the narrow-line regions in two nearby QSO2s. We construct dynamical models based on detailed photoionisation models of the emission-line gas, including the effects of internal dust, to apply to observations of large-scale outflows from these AGNs. We use Mrk 477 and Mrk 34 in order to test our models against recent \textit{HST} STIS observations of [O~III] emission-line kinematics, since these AGNs possess more energetic outflows than found in Seyfert galaxies.  We find that the outflows within 500 pc are consistent with radiative acceleration of dusty gas, however the outflows in Mrk 34 are significantly more extended and may not be directly accelerated by radiation. We characterise the properties of X-ray winds found from the expansion of [O~III]-emitting gas close to the black hole. We show that such winds possess the kinetic energy density to disturb [O~III] gas at $\sim$ 1.8 kpc, and have sufficient energy to entrain the [O~III] clouds at $\sim$ 1.2 kpc. Assuming that the X-ray wind possesses the same radial mass distribution as the [O~III] gas, we find that the peak kinetic luminosity for this wind is 2\% of Mrk 34's bolometric luminosity, which is in the 0.5\% - 5\% range required by some models for efficient feedback. Our work shows that, although the kinetic luminosity as measured from [O~III]-emitting gas is frequently low, X-ray winds may provide more than one order of magnitude higher kinetic power.
\end{abstract}



\section{Introduction}
Supermassive black holes (SMBH) can be found within the center of all massive galaxies \citep{magorrian1998a, kormendy2013a}. A small percentage of these are actively accreting material from the surrounding accretion disk, which we define as active galactic nuclei (AGN). This process of fueling the AGN and the subsequent feedback are widely acknowledged to play a critical role in the evolution of galaxies by expelling gas from the central regions of galaxies, shutting down their global star formation and regulating their stellar mass and size growth \citep[e.g.,][]{dave2012a, fabian2012a, vogelsberger2013a, heckman2014a, king2015a, naab2017a}.\par 

Evidence for AGN outflows, which can produce feedback, can be observed in emission in optical and infrared spectroscopy as high-velocity gas (up to  $\sim$ 7000 ${\rm km~s^{-1}}$ in the most extreme cases; \citet{perrotta2019a}), with full width at half maximum (FWHM) $>$ 250 ${\rm km~s^{-1}}$, inside the Narrow-Line Region (NLR), a region with typical electron density in the order of $10^{2} - 10^{6}$ ${\rm cm^{-3}}$, which often has a biconical morphology \citep{pogge1988a, pogge1988b}, where the ionised gas extends from the torus to distances between tens and thousands of pcs from their nucleus. AGN outflows can also be observed in absorption in the UV and X-ray spectra of Type 1 AGN \citep[e.g.,][]{laha2021a}. More energetic types of outflows can be found in broad absorption line (BAL) QSOs \citep[e.g.,][]{hazard1984a, knigge2008a}, which exhibit broad blueshifted absorption lines, with low-velocity edges at $\sim$ 0.007c – 0.03c and high-velocity edges at
$\sim$ 0.01c – 0.25c \citep{gibson2009a}. The minimum velocity for BALs reaches 0 ${\rm km~s^{-1}}$ and, in some cases, $>$0 ${\rm km~s^{-1}}$ \citep[e.g.,][]{mcgraw2017a}. The maximum centroid velocity in the rest wavelength interval between 1549\AA~and 1400\AA \citep[e.g.,][]{gibson2009a} can reach blueshifted velocities with magnitude greater than 0.06c. However, there are also documented cases of UV BALs reaching blueshifted velocities with magnitude of almost 0.2c \citep[e.g.,][]{rodriguezhidalgo2011a, hamann2013a}. In addition, the fastest X-ray outflows, or Ultra Fast Outflows (UFOs), are seen in a large fraction of local AGNs \citep{tombesi2010a,tombesi2010b, tombesi2015a}, and are detected via blueshifted X-ray absorption lines with velocities at $\sim$ 0.02c – 0.25c.  \par 

High-velocity clouds in the NLR have been attributed to outflows with mass outflow rates in the order of 1–10 ${\rm M\textsubscript{\(\odot\)}}$  ${\rm yr^{-1}}$ \citep[e.g.,][]{dibai1965a, crenshaw2009a,crenshaw2015a, storchi2010a, muller2011a}. The outflow rates exceed the mass accretion rates required for the observed AGN bolometric luminosities, which suggests that most of the fueling flow is blown out by radiation pressure and/or highly ionised winds \citep{everett2007a}. \par

Different mechanisms have been proposed to drive the outflowing gas \citep[e.g.,][]{mathews1974a, veilleux2005a, king2015a}. One possibility, given the strong radiation field in AGNs, is the radiative acceleration of outflows \citep[e.g.,][]{mckee1975a, shields1977a, icke1977a, murray1995a, chiang1996a, proga2000a, chelouche2001a, chelouche2003a, chelouche2003b, proga2004a}. Magnetic fields are also considered a candidate for launching winds near the accretion disk in AGNs \citep[e.g.,][]{blandford1982a, contopoulos1994a, bottorff1997a, bottorff2000a, bottorff2000b}. In addition, there are studies which have combined these two mechanisms, focusing on the importance of radiative acceleration within magnetic winds \citep[e.g.,][]{dekool1995a, everett2002a, everett2005a, mizumoto2019a}. A third proposed mechanism is thermal outflows. \citet{chelouche2005a} and \citet{everett2007a} studied the possibility of thermally driven (Parker) winds and their "ability" to explain some observed low-velocity ($v$ $\sim$ 500 ${\rm km~s^{-1}}$) X-ray absorption features. \par 

When it comes to radiative acceleration, the gas will be dust free at distances close to the source, i.e., within the sublimation radius \citep{barvainis1987a}, but may still be radiatively accelerated via bound-bound and bound-free transitions within the irradiated gas. For example, \citet{arav1994a} and \citet{arav1994b} modeled radiative acceleration of BAL clouds based on models of winds from hot stars \citep{castor1975a}. \citet{arav1994b} assumed that the confinement is nonthermal, and they found that they could achieve high radial velocities and realistic absorption line profiles, including the drop in opacity toward higher velocities, if the BAL gas originated within $\sim$ 0.1 pc of the central source and was dynamically coupled to a hot confining medium. \par 

Further from the source, e.g., in the NLR, along with bound-bound and bound-free transitions (see section \ref{sec:force_multiplier}), internal dust can provide a major source of opacity \citep{dopita2002a}. In regions where the dust is optically thin and dynamically well-coupled to the gas, the radiation pressure force on the dust plus gas can easily exceed the gravitational force, even for highly sub-Eddington luminosities \citep[e.g.,][]{phinney1989a, pier1992a, laor1993a}. Studies have been done on the effects of dust within the NLR \citep{netzer1993a} and various studies \citep[e.g.,][]{baldwin1991a,pier1995a, dopita2002a} have used photoionisation models to study how the radiation pressure exerted by the AGN acts upon dust within the ionised gas. Dust plays an important role in this process, as well as being responsible for producing photoelectric heating in the photoionised plasma and absorbing the ionising photons. \par

Connecting the inner wind with large-scale molecular outflows is crucial for obtaining a global view of outflows \citep{tombesi2015a}. One of the ways of exploring this connection is via studies of the NLR.
The first high-spatially resolved spectra of the NLRs were obtained as a product of the high-resolution observations obtained with the \textit{Hubble} Space Telescope (\textit{HST}) \citep[e.g.,][]{crenshaw2000a, crenshaw2000b, kraemer2000a, kraemer2000b, kaiser2000a}. More recently, there have been studies using ground-based Integral Field Units (IFU) observations \citep[e.g.,][]{barbosa2006a,storchi2010a, fischer2017a, gnilka2020a}. There have been several analyses of the outflow kinematics \citep[e.g.,][]{das2005a, das2006a, crenshaw2010a, fischer2010a, fischer2011a, fischer2013a, revalski2021a} in nearby AGN. These authors have found that the observed velocity pattern often possessed a signature of radial acceleration followed by deceleration.\par

\citet{crenshaw2015a} showed that the outflow rate in the NLR of the Seyfert galaxy NGC 4151 was much higher than expected from a nuclear outflow. This indicates that most of the outflowing ionised gas they observed must have originated at large distances from the SMBH, i.e., accelerated in situ\footnote{In this study, our kinematic models are more complex than the bi-conical model presented by \citet{fischer2013a}. }. One possibility is that the outflowing gas was accelerated off a large reservoir of gas that was accrued before the AGN "turned on". Support for this theory comes from the Gemini Near Infrared Field Spectrograph (NIFS) observations of another Seyfert galaxy, Mrk 573, by \citet{fischer2017a}. They observed evidence for outflows accelerated off circumnuclear dust spirals that cross into the NLR bicone. \par 

The Space Telescope Imaging Spectrograph (STIS) allows high spatial and spectral resolution studies of the NLR, offering data sufficient to constrain photoionisation and dynamical models of the inner region of AGNs. In particular, \citet{fischer2018a} measured the [O~III] 5007~\AA~velocities and line profile widths as a function of radial distance of 12 of the 15 brightest targets at z $\leq$ 0.12 from the \citet{reyes2008a} sample of QSO2s. The study characterized AGN-driven outflows in each QSO2 of the sample by analyzing the [O~III] morphology and emission-line kinematics of each target. QSOs were chosen for this study because of the luminosity effect, since they are more luminous than nearby AGN, i.e., Seyfert galaxies, that had been studied so far. More specifically, Type 2 quasars were chosen over Type 1s, since the observed NLR morphologies in QSO1s can be strongly affected by projection effects. \par 

Based on their kinematics, \citet{fischer2018a} categorized the influence of the central AGNs in three different regions, as a function of distance from the nucleus. In the inner region, the emission lines have multiple components with high central velocities and high FWHM, which are consistent with outflows. At greater distances, gas is still being ionised by the AGN radiation but emission lines exhibit low central velocities with low FWHM, consistent with rotation of the host galaxy. In addition, \citet{fischer2018a} identified a third kinematic component, at intermediate distances, which presents low central velocities but high FWHM. This kinematics is referred to as "disturbed" kinematics and they suggest that AGN activity may be disrupting this gas, causing large turbulent motion within the [O~III]-emitting gas\footnote{By [O~III]-emitting gas, we are referring to the state in which the ${\rm O^{++}}$ is the dominant ionisation state of oxygen. At lower or higher ionisation parameters, where there is less ${\rm O^{++}}$, the [O~III] emission becomes relatively weaker and other emissions from other ionisation states start to dominate. } without resulting in radial acceleration.\par

We have recently continued the analysis of these AGNs \citep[][hereafter Paper I]{trindadefalcao2020a}. Using the same data as \citet{fischer2018a}, we computed masses, mass outflow rates, kinetic energies, kinetic energy rates, momenta and momentum flow rates of these outflows. We concluded that the outflows in our sample contain a maximum total ionised gas mass of $3.4\times 10^{7}{\rm M\textsubscript{\(\odot\)}}$, and a maximum outflow rate of $10.3~{\rm M\textsubscript{\(\odot\)}}$ ${\rm yr^{-1}}$, both for Mrk 34. The ratios between the maximum kinetic luminosity and the bolometric luminosity for the entire sample, i.e., $3.4\times10^{-8}$ - $5\times10^{-4}$, indicate that the [O~III] winds are not an efficient feedback mechanism, based on the criteria of \citet{dimatteo2005a} and \citet{hopkins2010a}.\par 

In our current study, we use the same data as \citetalias{trindadefalcao2020a} and investigate the dynamics of the [O~III]-emitting gas in the NLR of two of the nearest QSO2s in our sample, Mrk 477 and Mrk 34, focusing on whether the gas can be radiatively accelerated. We also explore the idea of outflowing gas being accelerated in situ, which is consistent with the NGC 4151 and Mrk 573 analyses \citep{crenshaw2015a, fischer2017a}. Specifically for Mrk 34, we study the possibility that X-ray winds, which were radiatively accelerated while in a lower ionisation state, are responsible for the entrainment of the [O~III] gas at $\sim$ 1.2 kpc and the disturbed kinematics detected at $\sim$ 1.8 kpc, which we discuss in sections \ref{sec:entrainment} and \ref{sec:kinetic_energy}, respectively.

\section{Analysis of the Observations}
\label{sec:new_Observations}
In this section, we summarize the observations and subsequential analysis performed in this paper. The complete and detailed analysis of the \textit{HST} ACS and STIS data are presented in the studies by \citet{fischer2018a} and in \citetalias{trindadefalcao2020a}. \par

For this study, we select Mrk 477 and Mrk 34, two of the targets from our sample of QSO2s from \citetalias{trindadefalcao2020a}, to extend our analysis of the dynamics of the [O~III] gas. These AGNs have a redshift of $z$=0.038 and $z$=0.051, and a bolometric luminosity of $L_{bol} = 1.8\times 10^{45}$ ${\rm erg~s^{-1}}$ and $L_{bol} = 2.6\times 10^{45}$ ${\rm erg~s^{-1}}$, respectively (see  \citetalias[see][section 2.4]{trindadefalcao2020a}). Among the targets in the \citet{reyes2008a} sample, Mrk 34 and Mrk 477 are the two nearest QSO2s with the best-resolved NLRs and, hence, the targets for which the radial mass profile can most readily be constructed.

\subsection{The Radial Mass Profiles of Mrk 477 and Mrk 34}
\label{sub:radial_mass_profile}
We calculate the radial mass profiles of Mrk 477 and Mrk 34, which we use as inputs to our dynamical analysis described in Section \ref{sec:dynamics_outflows}.\par 

\begin{figure*}
  \centering
 \begin{minipage}[b]{0.3\textwidth}
  \includegraphics[width=5.8 cm, height=3.85cm]{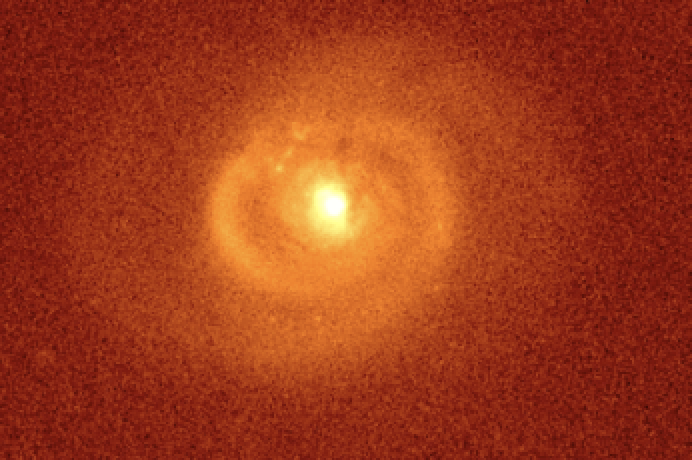}
 \end{minipage}\qquad
 \begin{minipage}[b]{0.3\textwidth}
  \includegraphics[width=5.8 cm, height=3.85cm]{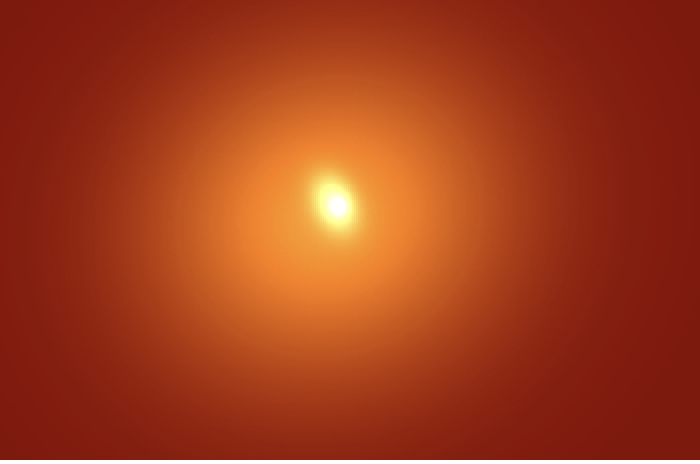}
 \end{minipage}\qquad
 \begin{minipage}[b]{0.3\textwidth}
  \includegraphics[width=5.8 cm, height=3.85cm]{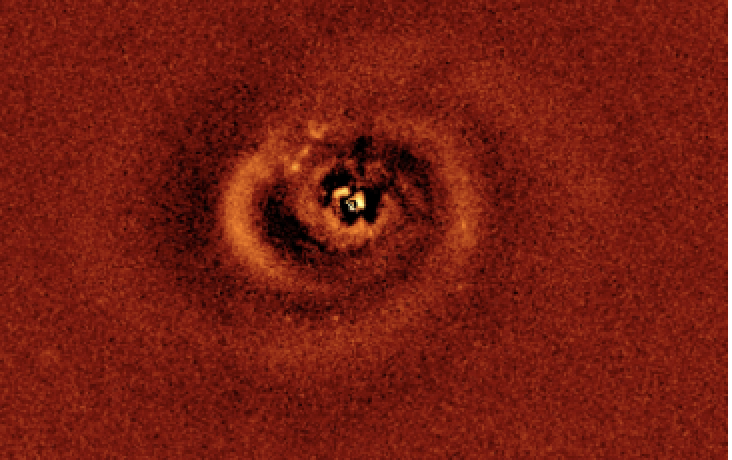}
 \end{minipage}\qquad
 \caption{Left panel: \textit{HST} WFPC3/PC F814W continuum image of Mrk 477. Middle panel: best fit galaxy decomposition model (3 components) for Mrk 477. Right panel: residuals between image and model.}
\label{fig:mrk477_mass}
\end{figure*}

\begin{figure*}
  \centering
 \begin{minipage}[b]{0.3\textwidth}
  \includegraphics[width=5.8 cm, height=3.85cm]{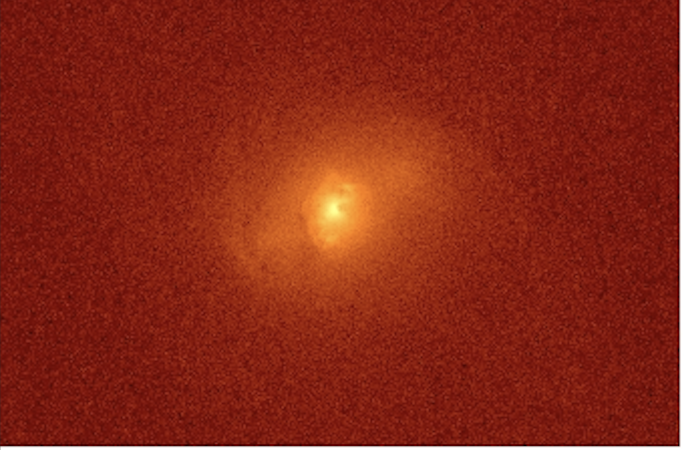}
 \end{minipage}\qquad
 \begin{minipage}[b]{0.3\textwidth}
  \includegraphics[width=5.8 cm, height=3.85cm]{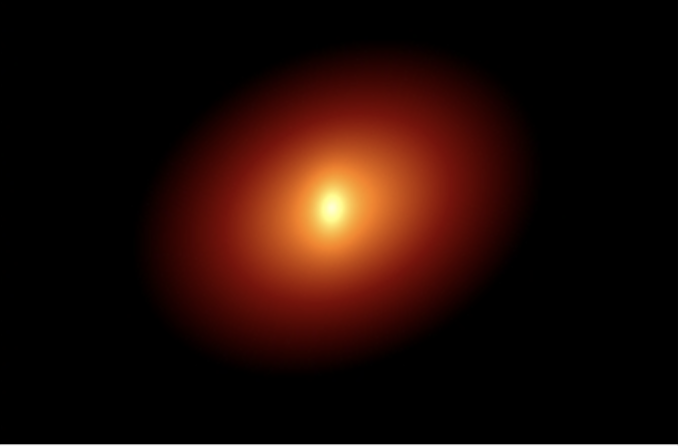}
 \end{minipage}\qquad
 \begin{minipage}[b]{0.3\textwidth}
  \includegraphics[width=5.8 cm, height=3.85cm]{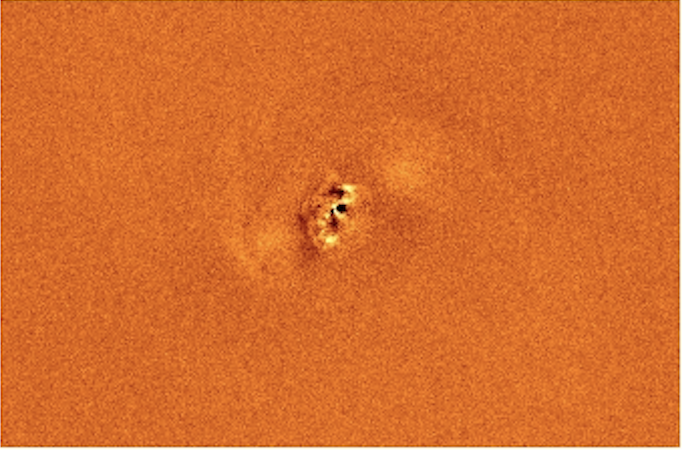}
 \end{minipage}\qquad
 \caption{Left panel: \textit{HST} WFPC3/PC F814W continuum image of Mrk 34. Middle panel: best fit galaxy decomposition model (2 components) for Mrk 34. Right panel: residuals between image and model.}
\label{fig:mark34_mass}
\end{figure*}

The radial mass distributions of Mrk 477 and Mrk 34 are determined using HST/WFPC3 F814W images, employing the method described by \citet{fischer2019a}; specifically, we use GALFIT version 3.0.5 \citep{peng2002a, peng2010a} to perform image decomposition to separate the galaxy/stellar light from the nuclear emission. We find that the best fitting model is composed of two Sérsic components for Mrk 34 and three components for Mrk 477. The best fitting parameters are presented in Table \ref{tab:Table_1}, and the original images, GALFIT models, and resulting residual maps, are presented in Figures \ref{fig:mrk477_mass} and \ref{fig:mark34_mass} for Mrk 477 and Mrk 34, respectively. In the case of Mrk 477 we find two inner components that can be identified with a pseudo-bulge and a bar, with the addition of a third component that corresponds to the disk. In the case of Mrk 34 the two components correspond to a pseudo-bulge and a disk. \par

\begin{figure*}
  \centering
 \begin{minipage}[b]{0.45\columnwidth}
  \advance\leftskip-5cm
  \includegraphics[width=9cm, height=6cm]{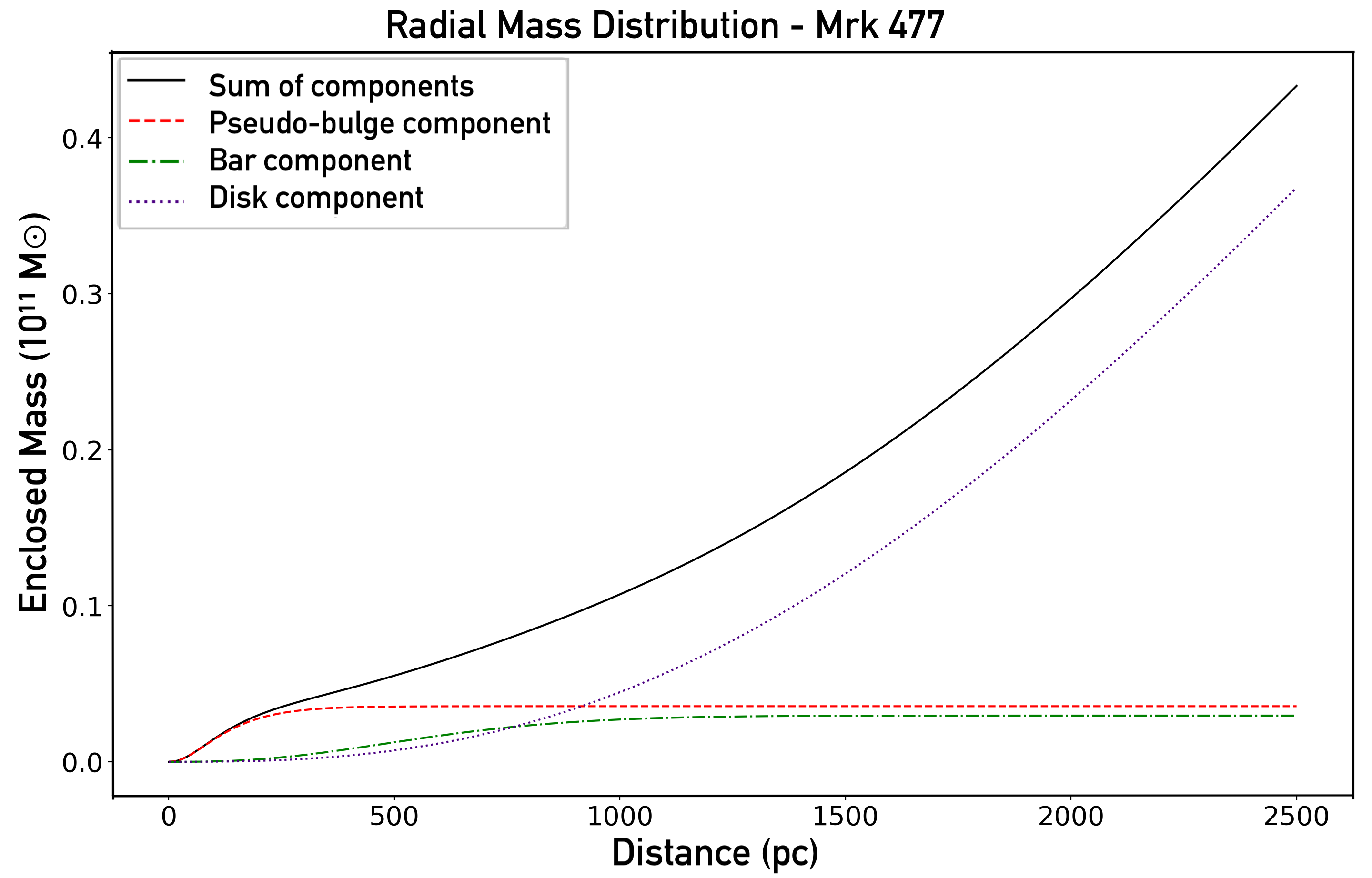}
 \end{minipage}\qquad
 \begin{minipage}[b]{0.45\columnwidth}
  \includegraphics[width=9.2cm, height=6.1cm]{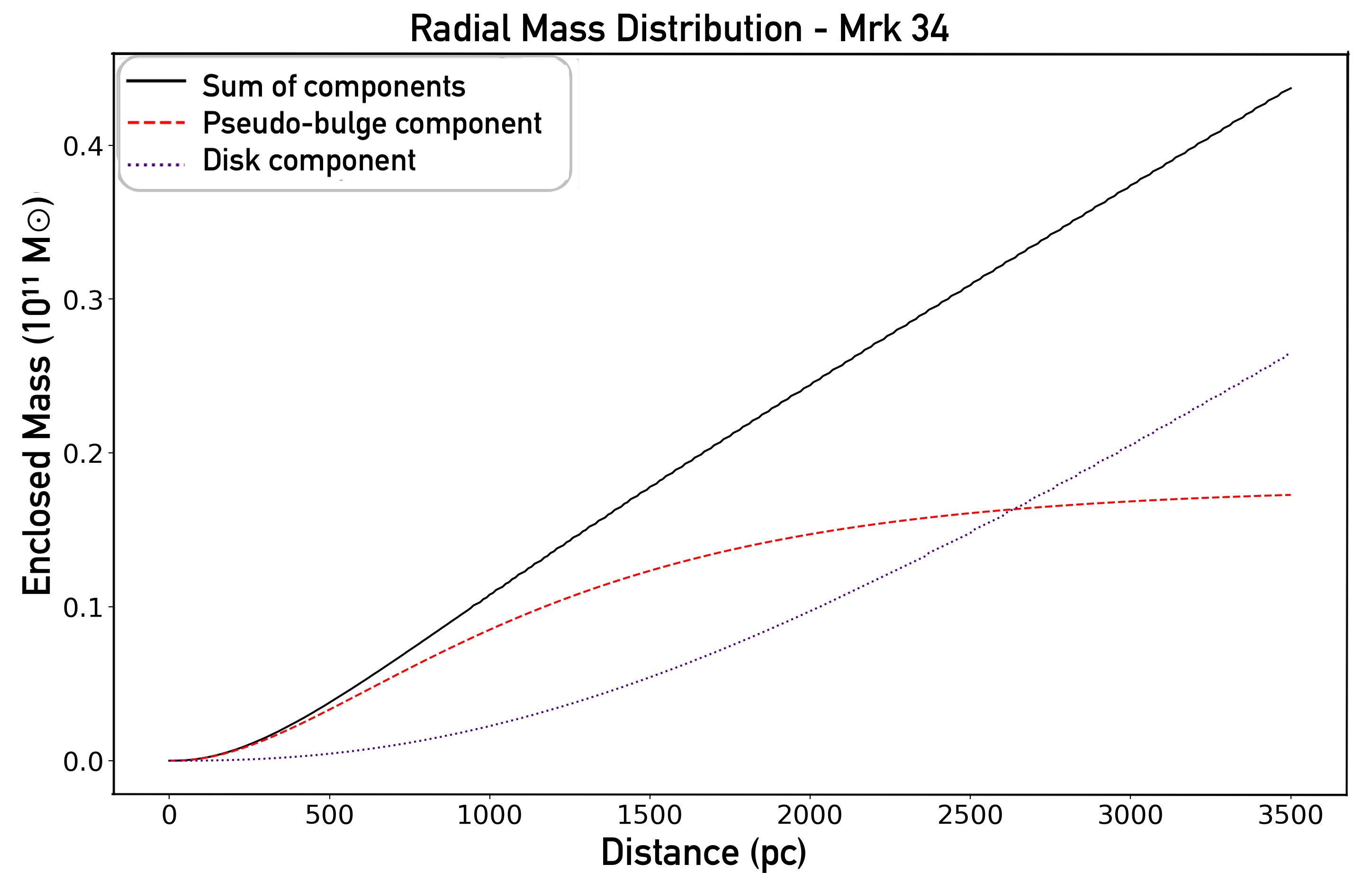}
 \end{minipage}\qquad
 \caption{Radial mass distribution profiles for each component in our model for Mrk 477 (left panel) and Mrk 34 (right panel). Green dashed-dotted, red dashed, purple dotted and black lines represent the bar component, the pseudo-bulge component, the disk component and the sum of all the components, respectively. Our radial mass distribution is calculated using the expressions from \citet{terzic2005a} assuming a mass-to-light ratio of 1.2 for Mrk 477, and a ratio of 2 for Mrk 34, as described in section \ref{sub:radial_mass_profile}. }
\label{fig:radial_mass}
\end{figure*}

The radial mass distributions of the individual Sérsic components, as well as the total radial mass distributions are presented in Figure \ref{fig:radial_mass}, for Mrk 477 and Mrk 34. They were calculated using the GALFIT results and the expressions from \citet{terzic2005a}, assuming a distance of 173 Mpc for Mrk 477 and 218 Mpc for Mrk 34. Given that the stellar population of Mrk34 is dominated by old stars \citep{raimann2003a}, we assume a mass-to-light ratio of 2 for this galaxy, which is consistent with the F814W filter used for the observations. We also assume I$_{F814W}=4.12$ mag for the absolute magnitude of the Sun in the Vega system.\par 

\begin{table}
\caption{GALFIT model results.}
\label{tab:Table_1}
\begin{tabular}[t]{lccccc}
\toprule

 Component & I$_{F814W}$ & ${\rm r_{e}}$  & n & b/a & PA \\
 & (mag) & (kpc) & & (deg)\\
\midrule
& &  \textbf{Mrk 34} \\
pseudo-bulge & 15.94 & 0.78 & 1.27 & 0.71 & 177\\
disk & 14.07 & 4.13 & 1.04 & 0.74 & 116\\
\midrule
& &  \textbf{Mrk 477} \\
pseudo-bulge & 16.63 & 0.09 & 1.13 & 0.89 & 6\\
bar & 16.83 & 0.42 & 0.62 & 0.69 & 25\\
disk & 14.30 & 2.81 & 0.70 & 0.89 & 110\\
\bottomrule
\end{tabular}
\end{table}

In the case of Mrk 477, the mass-to-light ratio is smaller, given that this galaxy is known to harbor a circumnuclear starburst \citep{heckman1997a}. In order to determine a more appropriate mass-to-light ratio, we scale the value used for Mrk 34 based on the spatially resolved stellar population synthesis results from \citet{raimann2003a}. An inspection of their results (see their Figures 8 and 16, as well as Table 4) indicates that in the cases of Mrk 477 and Mrk 34 there is not a large variation in the stellar population as a function of radial distance up to $\sim$5 kpc from the nucleus. Given that \citet{raimann2003a} present the fractional contribution of the different stellar population age components to the light at 4020~\AA, we convert these values to 8040~\AA\ using information about the continuum properties of these components from \citet{bica1986a} and \citet{bica1987a}. We find that the stellar population of Mrk 34 resembles that of an S6 template \citep{bica1988b}, composed mostly of old stars, with a small contribution (on the order of $\sim$10\%) from stars with ages of 100 Myr and younger to the light at 5870~\AA. In the case of Mrk 477, the stellar population resembles that of an S7 template, with similar contributions from old stars and stars with ages of 100 Myr and younger, to the light at 5870~\AA. \par

Combining these results with those for the mass-to-light ratio of the different age components at 8040~\AA \citep{bica1988a}, we determine the mass fraction from each age component to the spectrum of Mrk 34 and Mrk 477, as well as the summed value at different distances from the nucleus. We find that there is a variation as function of distance from the nucleus that is smaller than 10\%. The mass-to-light ratio of Mrk 477 corresponds to $\sim$0.6 of that of Mrk 34. Based on these results we use a mass-to-light ratio of 1.2 for Mrk 477 in the F814W band. Note that Mrk 477 has a small bulge but a large disk, which dominates its radial mass profile starting at $\sim$ 700 pc (see the purple dotted line in the left panel of Figure \ref{fig:radial_mass}). Meanwhile, Mrk 34 possesses a bigger bulge that dominates the radial mass profile as far as $\sim$ 2.5 kpc (see the red dashed line in the right panel of Figure \ref{fig:radial_mass}).\par 

It should be noted that a similar light profile decomposition done by \citet{zhao2019a}, on these same data, obtained structures with different parameters. Some of the differences between their results and ours are due to the fact that they adopt an exponential disk profile (Sérsic n$=1$) and the use of a PSF model. However, we find that their integrated mass distributions and radial mass profiles do not differ significantly from ours. In the case of Mrk 34, both galaxy models have the same mass, although \citet{zhao2019a} find a bulge mass 2.5 times larger than ours. A more detailed inspection shows that their model results in a more centrally peaked radial mass profile, with 2.5 times more mass in the inner 100 pc, reducing to a difference of $\sim$10 percent at 500 pc and less than 1 percent at distance larger than 1 kpc. In the case of Mrk 477, \citet{zhao2019a} find a bulge mass 50 percent larger than that of our 2 inner components, while our integrated mass is 2.5 times larger than theirs. The higher integrated mass can be attributed to our choice of mass-to-light ratio for the disk. Since the velocity of the gas is proportional to $\sqrt{M(r)}$ (see Equation 3), this difference in the integrated mass of these galaxies do not alter the overall results of the paper. 

\subsection{\textit{Chandra} Imaging Analysis of Mrk 34}
\label{sub:chandra_imaging}
The \textit{Chandra}/Advanced CCD Imaging Spectrometer (ACIS) observation of Mrk 34, with an exposure time of 100 ksec and a band-pass of 0.3 keV - 10 keV, was obtained on 30-Jan-2017 (Obsid 18121, PI:Elvis).  Unless otherwise specified, calculations are obtained using DS9, standard CIAO software tools \citep{fruscione2006a} and \citep{gehrels1986a} uncertainties.  Pileup was negligible;  the brightest pixel in the native-binned event file has 134 counts in 100 ksec.  We use sub-pixel imaging at 1/8th scale and convert the count rates to flux using the webPIMMS\footnote{https://cxc.harvard.edu/toolkit/pimms.jsp} software tool and Galactic $N_{H}$ column density determined from COLDEN\footnote{https://cxc.harvard.edu/toolkit/colden.jsp} using NRAO maps \citep{dickey1990a}.  The relevant energy bands are sufficiently narrow that our flux values are insensitive to choice of model, and the maximum off-axis separation in the region of interest is small enough that it does not introduce effects that are significant relative to the Poission noise. We assume that the nucleus is located at the spatial centroid of FeK$\alpha$ (6-7 keV).  We extract flux from a $1\farcs62$ wide strip at the same angle as the \textit{HST} STIS G430M slit in \citet{revalski2018a}.  This strip is broken into $0\farcs25$ extraction bins in order to take advantage of the \textit{Chandra} angular resolution. The primary source of spatial uncertainty in the X-ray radial flux distribution is the ACIS PSF and source counting statistics. Using wavdetect\footnote{Wavdetect is a standard CIAO source detection tool for Chandra (https://cxc.cfa.harvard.edu/ciao/ahelp/wavdetect.html) and one of the most commonly-used one.  It correlates the image with Mexican hat waveforms at a range of different spatial scales and looks for significant correlations.  It also produces a catalog of source properties, which helps to produce a rigorous estimate of the positional uncertainty.} to estimate the uncertainty of the Fe Ka (6-7 keV) position we obtain $\sim$ $0\farcs04$.  We use CIAO and MARX to simulate a monoenergy PSF at the band midpoint to estimate resolution effects.  On this grid, the 1-dimensional width for a point source is $0\farcs310$ ($2\times0\farcs155$) for 50\% encircled energy, and $0\farcs496$ ($2\times0\farcs248$) for 68.27\% encircled energy ($1\sigma$).  Each bin would enclose 42\% of the counts for a point source in its center, and each adjoining bin contains 21\%.  Hence, it is likely that $\sim$ 50\% of the counts in any given bin come from adjoining bins, and that $\sim$ 50\% of the true counts in the low central bin could also have been redistributed to adjoining bins.  As we are in the Poisson uncertainty regime, we limit positive detections to 3-sigma features that span at least 2 bins.

\section{Radiatively Driven Dynamics}
\label{sec:radiatively_driven}
In order to determine whether the gas is radiatively accelerated, we solve the equation of motion for gas under the influence of gravity and radiation pressure force \citep[e.g.,][]{das2007a}. Assuming the acceleration is a function of radial distance, it can be written as:
\begin{equation}
    a(r) = \frac{L\sigma_{T}\mathcal{M}}{4\pi r^{2}c\mu m_{p}}-\frac{GM(r)}{r^{2}}
\end{equation}
\noindent where $L$ is the effective bolometric luminosity of the AGN (see section \ref{sec:force_multiplier}), $\sigma_{T}$ is the Thomson scattering cross section for the electron, $\mathcal{M}$ is the force multiplier (see section \ref{sec:force_multiplier}), $r$ is the distance from the SMBH, $c$ is the speed of light, $\mu$ is the mean mass per proton, which for our study we consider to be 1.4, $m_{p}$ is the mass of the proton, $G$ is the universal gravitational constant and $M(r)$ is the enclosed mass at the distance $r$ determined from the radial mass distribution (see section \ref{sub:radial_mass_profile}). \par 
Equation 1 can be rewritten in terms of the velocity as a function of the distance from the SMBH, as follows:
\begin{equation}
    v dv =  \frac{L\sigma_{T}\mathcal{M}}{4\pi r^{2}c\mu m_{p}} dr - \frac{GM(r)}{r^{2}} dr 
\end{equation}
Converting Equation 2 to units of ${\rm km~s^{-1}}$ and pc, then integrating and setting the initial velocity to zero, we can rewrite the velocity of the gas as:
\begin{equation}
    v(r) = \sqrt{\int_{r_1}^{r} [4886L_{44}\frac{\mathcal{M}}{r^2}-8.6\times10^{-3}\frac{M(r)}{r^2}]\, dr}~~\textrm{km~${\rm s^{-1}}$}
\end{equation}\par
\noindent where $L_{44}$ is the effective bolometric luminosity in units of $10^{44}$ ${\rm ergs~s^{-1}}$, $r_1$ is the launch radius of the gas, and $M(r)$ is expressed in solar masses. \par

We do not consider the deceleration of the clouds by drag forces, e.g., ram pressure, due to interaction with the interstellar medium of the host galaxy, which means that we may be overestimating our computed velocities. According to \citet{das2007a}, for an ambient density of 0.1 ${\rm cm^{-3}}$, there is a factor of 2 decrease in velocity. For higher densities, e.g., 1 ${\rm cm^{-3}}$, it is unlikely that these outflows could be launched at all (see \citealt{das2007a}, their Figure 4). \par 

We use photoionisation models generated with Cloudy \citep{ferland2017a}, which calculates conditions of the gas in zones starting at the side of the cloud closest to the source of ionisation radiation, to determine inputs for the dynamical calculation, such as $\mathcal{M}$. Therefore it is useful to briefly summarise the model inputs we used in \citetalias{trindadefalcao2020a}. The model spectral energy distribution (SED) is of the form:
 
    $L_{\nu} \propto \nu^{-\alpha}$

\noindent where $ \alpha$, the spectral or energy index, is a positive number \citep[e.g.,][]{laor1997a, melendez2011a}. We assume that the UV to low energy, “soft” X-ray SED, is characterised by one value of $ \alpha$, while the high energy, “hard” X-ray SED, has a lower value of $ \alpha$. For our study we adopt a cutoff at 100 keV and we set the breakpoint at 500 eV, using the following values \citep{revalski2018a}: 
\medskip

 $ \alpha$ = 1.0 for 1 eV $\leq$ $h \nu$ $<$ 13.6 eV;  \par
 $ \alpha$ = 1.4 for 13.6 eV $\leq$ ${ h \nu}$ $\leq$ 500 eV;  \par
 $ \alpha$ = 1.0 for 500 eV $\leq$ ${ h \nu}$ $\leq$ 10 keV; \par 
 $ \alpha$ = 0.5 for 10 keV $\leq$ ${ h \nu}$ $\leq$ 100 keV; \par
\medskip
\noindent We also assumed elemental abundances of 1.4 $Z_{sun}$ (see \citetalias{trindadefalcao2020a}).

\subsection{Force Multiplier}
\label{sec:force_multiplier}
The force multiplier ($\mathcal{M}$) is defined as the ratio of the total absorption and scattering cross-section of the accelerated dust and gas to the Thomson cross-sections \citep{castor1975a,abbott1982a,crenshaw2003a}. In addition to Compton scattering, the momentum transferred to the gas is a result of the absorption of continuum radiation by bound-bound, bound-free, Compton scattering and free-free processes. The importance of these various processes depends on the atomic cross-section, the ionisation state of the gas, which can be quantified in terms of the ionisation parameter\footnote{The ionisation parameter is given here by $U = \frac{Q}{4\pi cr^{2}n_{H}}$, where $Q$ is the number of ionising photons per second emitted by the AGN, $r$ is the radial distance from the AGN, $c$ is the speed of light and $n_{H}$ is the hydrogen number density.}, $U$, and the SED. \par 

The largest cross-sections occur for bound-bound transitions, which are observed as UV resonance lines. The most important ones for [O~III]-emitting gas (which we will refer to as [O~III]-gas) occur at energies lower than the ionisation potential of hydrogen, i.e., h$\nu$ $<$ 13.6 eV. For our assumed SED, the value of $L_{\nu}$ is larger than the ionisation potential of hydrogen value at the ionisation threshold. However, at large optical depth, large cross-sections imply that the gas becomes "self-shielded", which means that bound-bound absorption contributes the most at the illuminated surface, but decays quickly into the cloud (see \citealt{chelouche2001a}, their Figure 1)\footnote{The bound-bound absorption can be enhanced if there is internal turbulence. However, here we assume that the gas has thermal velocity dispersion.}. The bound-free transitions are the next largest contributing process to the total cross section . The cross-section for Compton scattering is equal to the Thomson cross-section, $\sigma_{T} = 6.65 \times 10^{-25}~{\rm cm^{-2}}$, since we are in the non-relativistic regime, therefore Compton scattering is negligible unless the column density is $\gtrsim$ $10^{24}~{\rm cm^{-2}}$. However, Compton scattering is the main source of opacity in highly ionised gas, e.g., with only H-like states remaining for the most abundant heavy elements, since the other sources of opacity, e.g., bound-bound and bound-free transition, are negligible in this regime. For most cases, free-free absorption contributes very little to the radiation pressure ($\sim$ 0.01 Compton). \par

\begin{figure*}
  \centering
 \begin{minipage}[b]{0.45\columnwidth}
  \advance\leftskip-5cm
  \includegraphics[width=9cm, height=6cm]{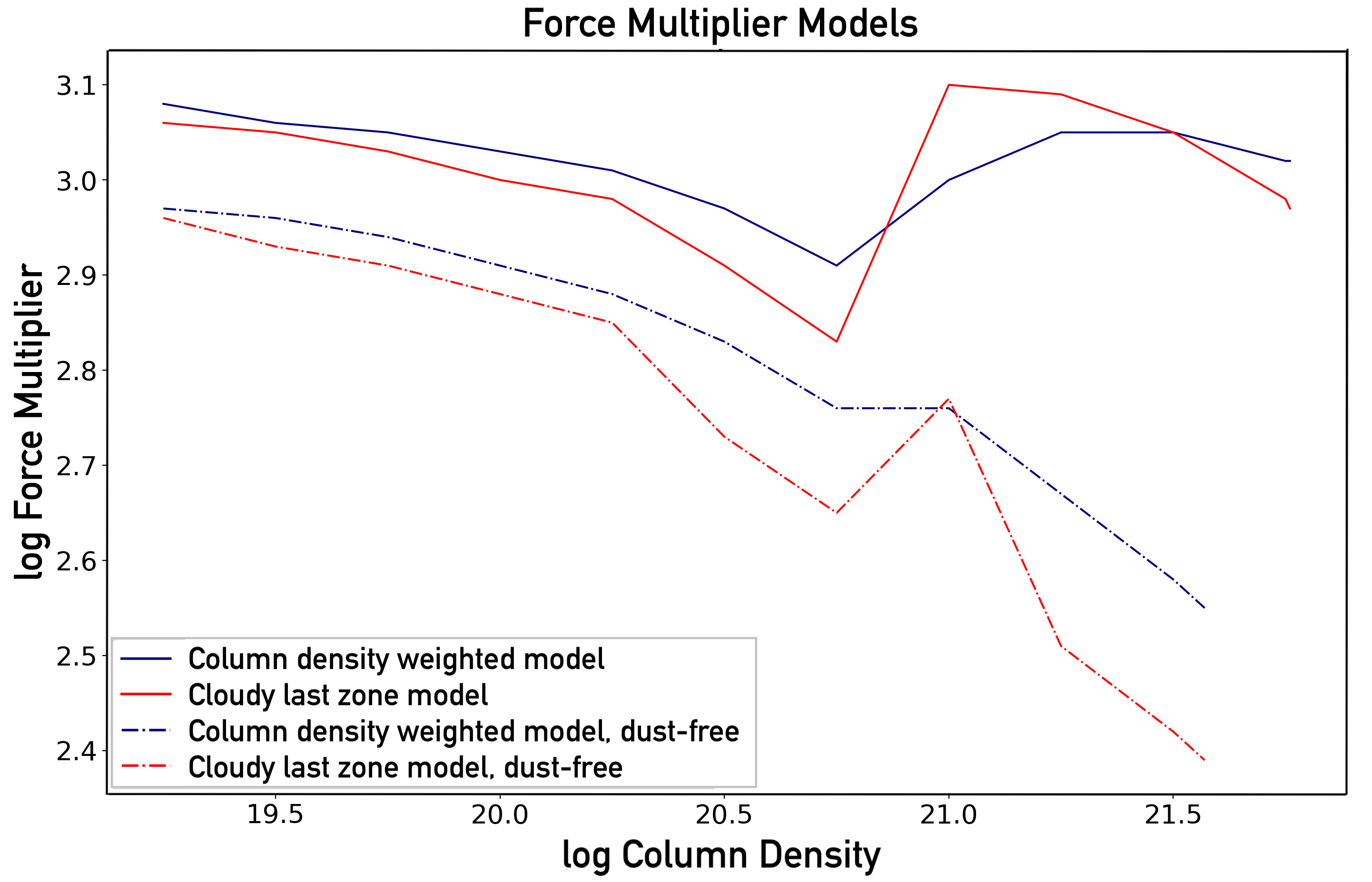}
 \end{minipage}\qquad
 \begin{minipage}[b]{0.45\columnwidth}
  \includegraphics[width=9.2cm, height=6.1cm]{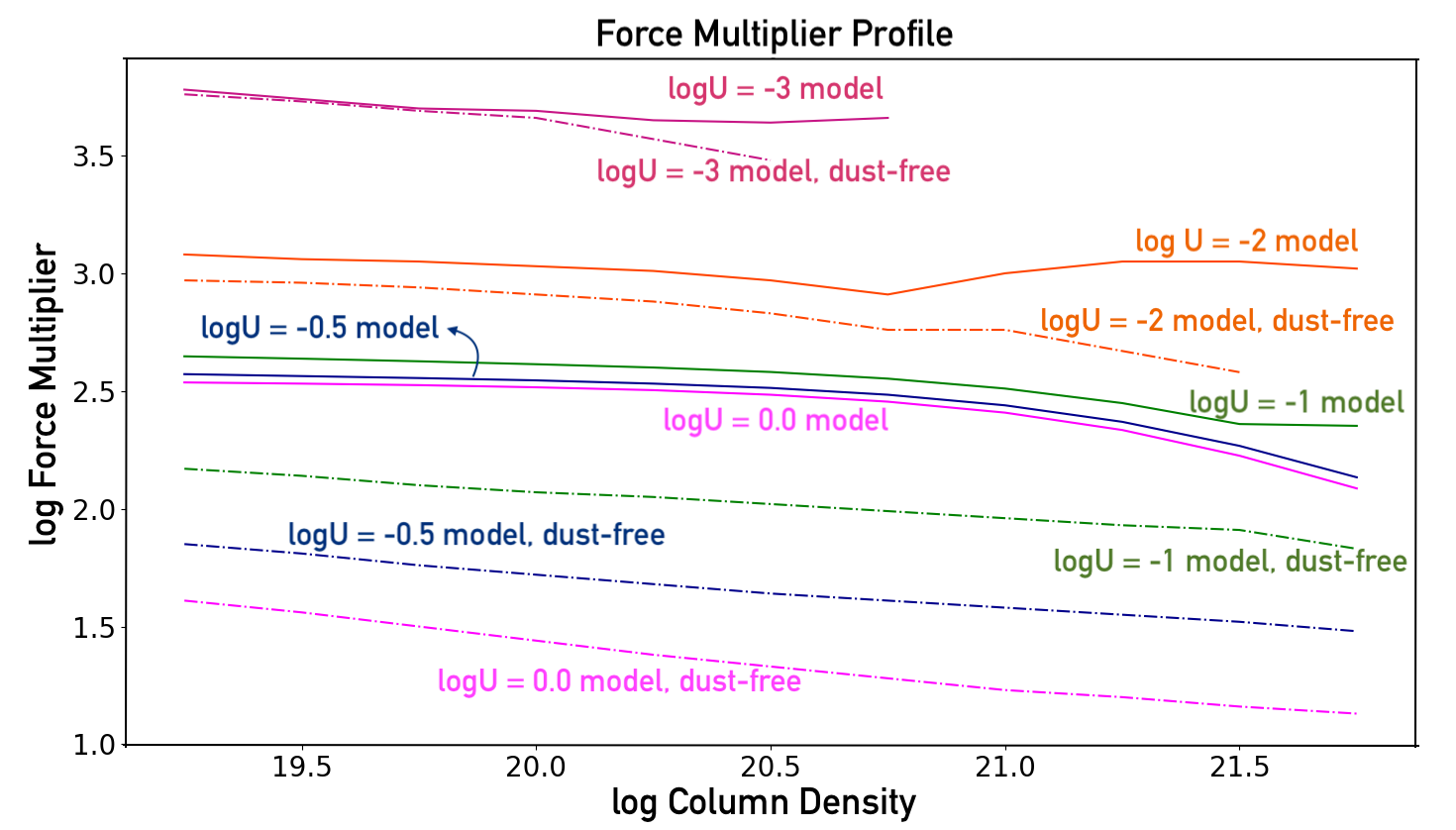}
 \end{minipage}\qquad
 \caption{Left panel: Comparison between the different methods for the calculation of the force multiplier as a function of depth into the cloud. All the models are for a gas with a logU =-2. There is an increase on the value of the force multiplier when it reaches the edge of a Strömgren sphere. This occurs because, as the hydrogen goes neutral, the total bound-free cross-section for hydrogen increases, with the result that the force multiplier value also increases. Right panel: Column density weighted force multiplier profile as a function of column density. The solid lines represent the dusty models and the dashed lines the dust-free models. The models are computed for constant density. The diagram shows how the value of $\mathcal{M}$ varies with different ionisation parameters and the changes that occur when internal dust is considered.}
\label{fig:force_multiplier}
\end{figure*}

The value of $\mathcal{M}$ at a given point in a gas cloud generally decreases as the column density and hence the optical depth increases. As the UV resonance lines saturate and the bound-free edges become optically thick, there is no contribution to the gas opacity at those particular wavelengths, which means that the effective cross-section per particle decreases, and, therefore, so does the $\mathcal{M}$ \citep[e.g.,][]{blumenthal1979a, mathews1982a,chelouche2001a}. $\mathcal{M}$ also depends on the ionisation state of the gas. As the ionisation parameter increases, the value for $\mathcal{M}$ decreases (see Figure \ref{fig:force_multiplier}), since there are fewer bound electrons available and the number of bound-bound and bound-free transitions decreases.\par

One can do a number of different things to estimate a more accurate $\mathcal{M}$ for the entire cloud. One possibility is to compute a column density ($N_{H}$) weighted $\mathcal{M}$, as follows: 
\begin{equation}
    \mathcal{M} = \frac{\sum \mathcal{M}_{i} (N_{H_{i}} - N_{H_{i-1}})}{N_{H_{f}}}
\end{equation}
\noindent where $\mathcal{M}$ is the column density weighted force multiplier, $\mathcal{M}_{i}$ is the force multiplier from the i-th zone of Cloudy, $N_{H_{i}}$ is the column density for the i-th zone inside the cloud, and $N_{H_{f}}$ is the column density where the integration stops for the last of the set of models. In most cases, the results are slightly larger than the final zone original values (except when it reaches the edge of a Strömgren sphere, since, as the hydrogen goes neutral, the total bound-free cross-section for hydrogen increases). This is because, by "weighting" $\mathcal{M}$ by the column density, the contributions of gas characterized by higher $\mathcal{M}$ are included in the calculation (see Figure \ref{fig:force_multiplier}). \par 

As noted in the introduction, another source of opacity is internal dust \citep{dopita2002a}. The value of $\mathcal{M}$ increases, particularly at higher column densities,  when dust is included in the models (see Figure \ref{fig:force_multiplier}). For all our models in this study, we assume grain abundances of 50\% of those determined for the interstellar medium (ISM) \citep[e.g.,][]{mathis1977a},
with the depletions from gas phase scaled accordingly \citep[e.g.,][]{snow1996a}. We also use elemental abundances of $Z/Z_{sun} = 1.4$ (see \citetalias{trindadefalcao2020a}).Assuming a dust-to-gas ratio equal to that of the ISM results in a computed value of $\mathcal{M}$ $\sim$ 1.3 times larger than the 50\% ISM models, at $N_{H} = 10^{21}~{\rm cm^{-2}}$.\par 

Unlike bound-free absorption, the dust cross-section is large at energies $<$ 13.6 eV, which provides a source of opacity after the bound-bound transitions saturate and the ionising radiation is absorbed. At very high ionisation, Compton scattering is the dominant source of opacity unless there is dust present, since the dust absorption cross-section is much larger than $\sigma_{T}$ \citep{king2015a}. The difference can be seen when the $\mathcal{M}$ values for the dusty and dust-free models with log$U$ $>$ -0.5 are compared (see Figure \ref{fig:force_multiplier}).

\begin{figure}
  \centering
  \includegraphics[width=0.5\textwidth]{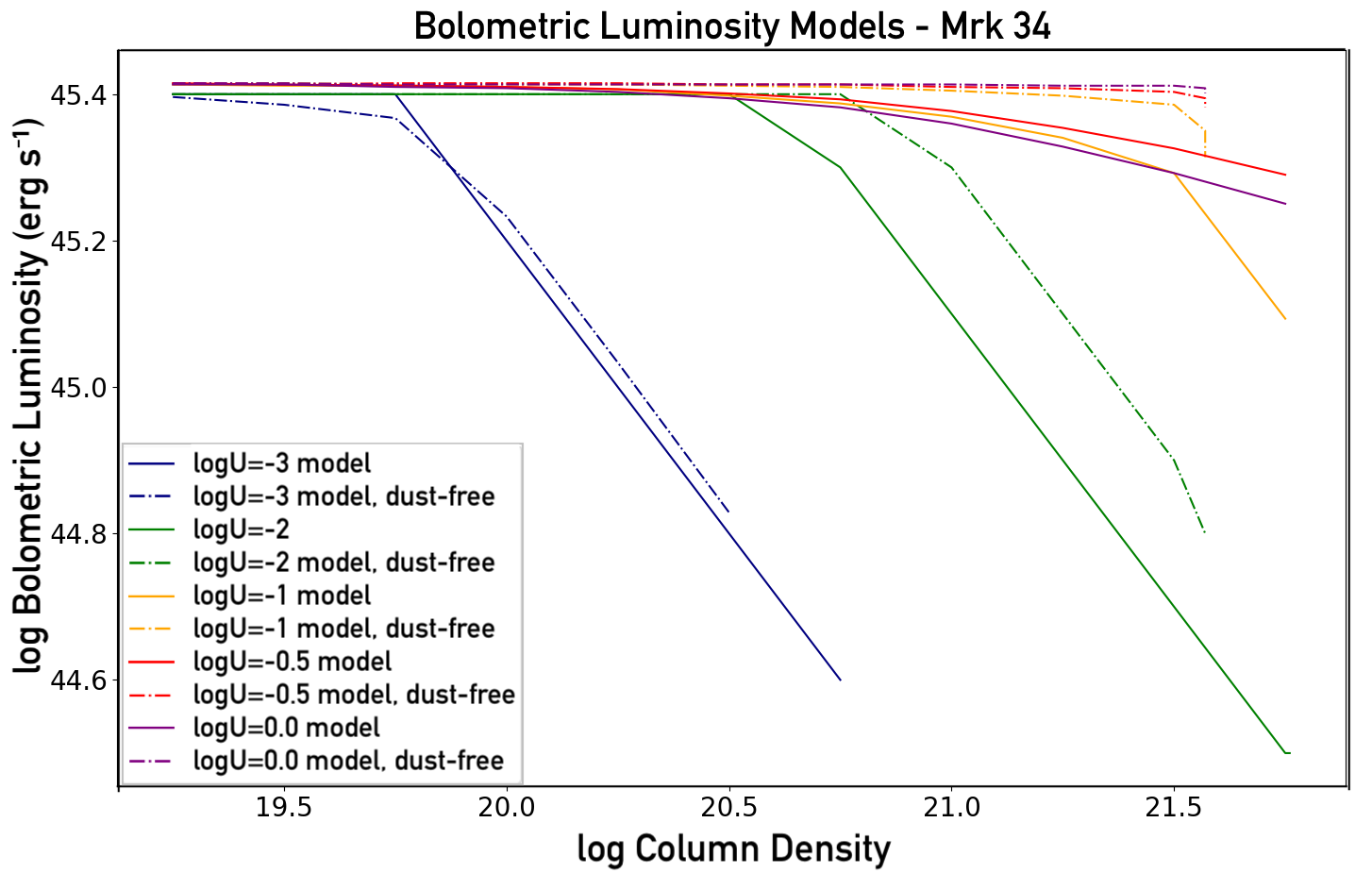}
\caption{\label{fig:bolometric_luminosity} Effective bolometric luminosity profile as a function of column density for Mrk 34. The solid lines represent the dusty models and the dashed lines the dust-free models. All models are computed for constant density.}
\end{figure}

A cloud\footnote{By using the term cloud we are not making any assumptions about the geometry of the [O~III]-emitting gas. Cloud is used in a descriptive way only.} of gas needs to be partially confined in order to be radiatively accelerated efficiently, because it has to stay long enough in a given state for this to happen \citep{blumenthal1979a}. A cloud can be confined by different means, such as ambient gas pressure, radiation pressure, and magnetic pressure. If gas is in the [O~III]-emitting state over its full trajectory, for example, the density of the gas must drop as $r^{-2}$ in order to maintain a constant $U$, which means that some sort of confinement is required (see section \ref{sec:expansion_rates}). On the other hand, if the cloud is not confined, it will expand freely, dropping its density and increasing its ionisation state with distance and, therefore, dropping its $\mathcal{M}$ (which is a strong function of the ionisation state, see Figure \ref{fig:force_multiplier}). Clouds will be confined in the radial direction by radiation pressure at the illuminated face or by ram pressure by the ambient medium, which will slow the rate of expansion \citep{blumenthal1979a, mathews1982a}. \par 

 In Equation 1, the acceleration of the gas depends on both the luminosity and the force multiplier. However, as noted by \citet{netzer2010a}, incident radiation is absorbed as the optical depth increases, which further reduces the amount of acceleration. In order to account for this, we compute an effective $L_{bol}$, based on the column density of the gas, as follows:
 
 \begin{equation}
    {L_{bol}} = \frac{\sum {L_{bol}}_{i} (N_{H_{i}} - N_{H_{i-1}})}{N_{H_{f}}}
\end{equation}

\noindent where ${L_{bol}}$ is the effective bolometric luminosity, in ${\rm ergs~s^{-1}}$, ${L_{bol}}_{i}$ is the calculated total luminosity after the i-th zone inside the cloud, $N_{H_{i}}$ is the column density for the i-th zone inside the cloud, and $N_{H_{f}}$ is the column density where the integration stops for the last of the set of models, i.e., the column density at which the gas temperature falls below 4000K. The effective bolometric luminosity profile is shown in Figure \ref{fig:bolometric_luminosity}. \par
 
 For comparison, \citet{netzer2010a} use an approximation for $\mathcal{M}$ that depends on the column density ($N_{H}$) and the fraction of radiation absorbed, i.e., $\mathcal{M}$ $\simeq$ $\alpha (r)/(\sigma_{T}N_{H})$, where $\alpha (r)$ is the fraction of the total bolometric luminosity absorbed by the cloud. Their method gives comparable values to our combination of weighted $\mathcal{M}$ and effective ${L_{bol}}$. \par

\begin{figure*}
  \centering
 \begin{minipage}[b]{0.45\columnwidth}
  \advance\leftskip-5cm
  \includegraphics[width=9cm, height=6cm]{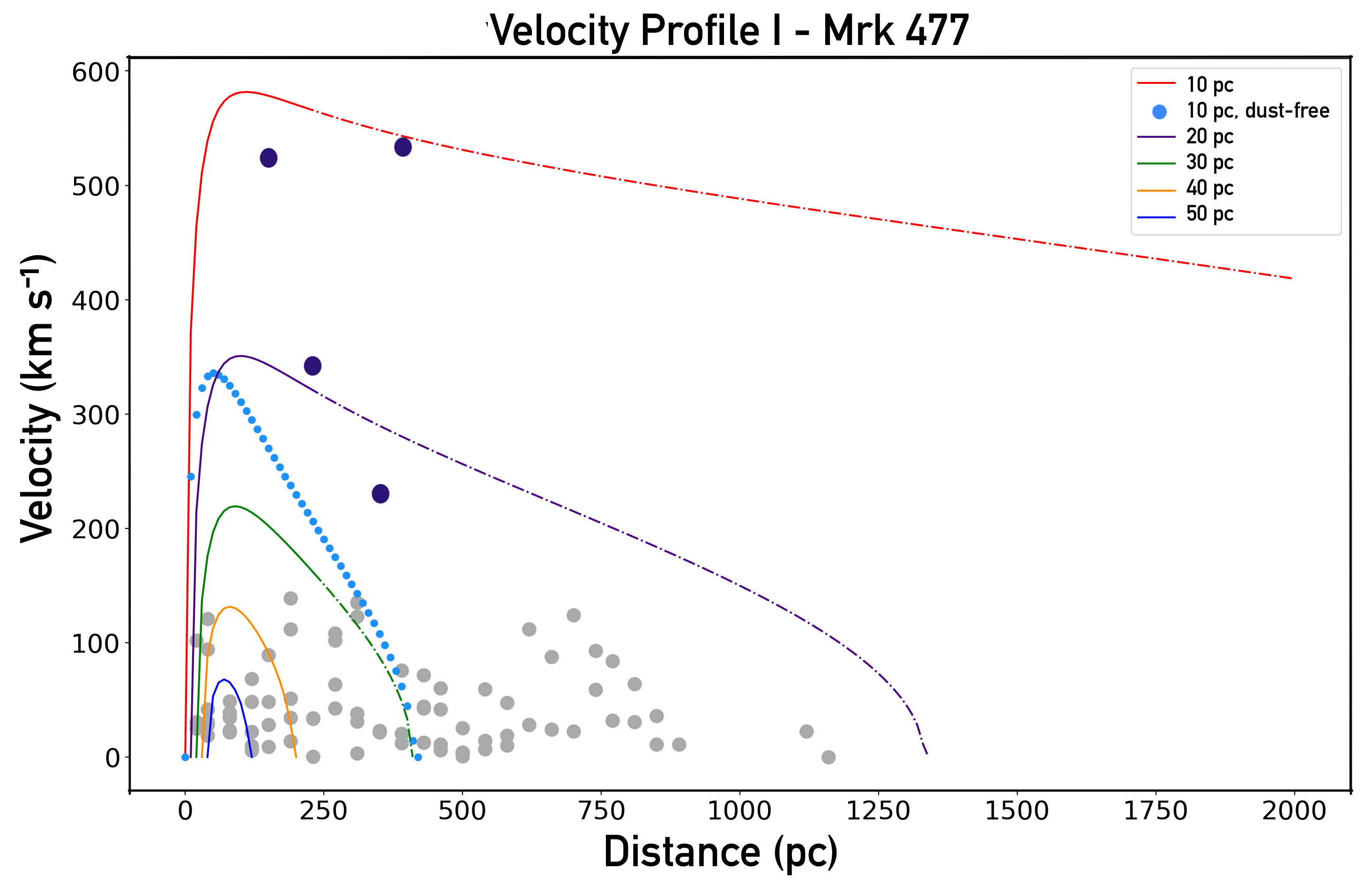}
 \end{minipage}\qquad
 \begin{minipage}[b]{0.45\columnwidth}
  \includegraphics[width=9cm, height=6cm]{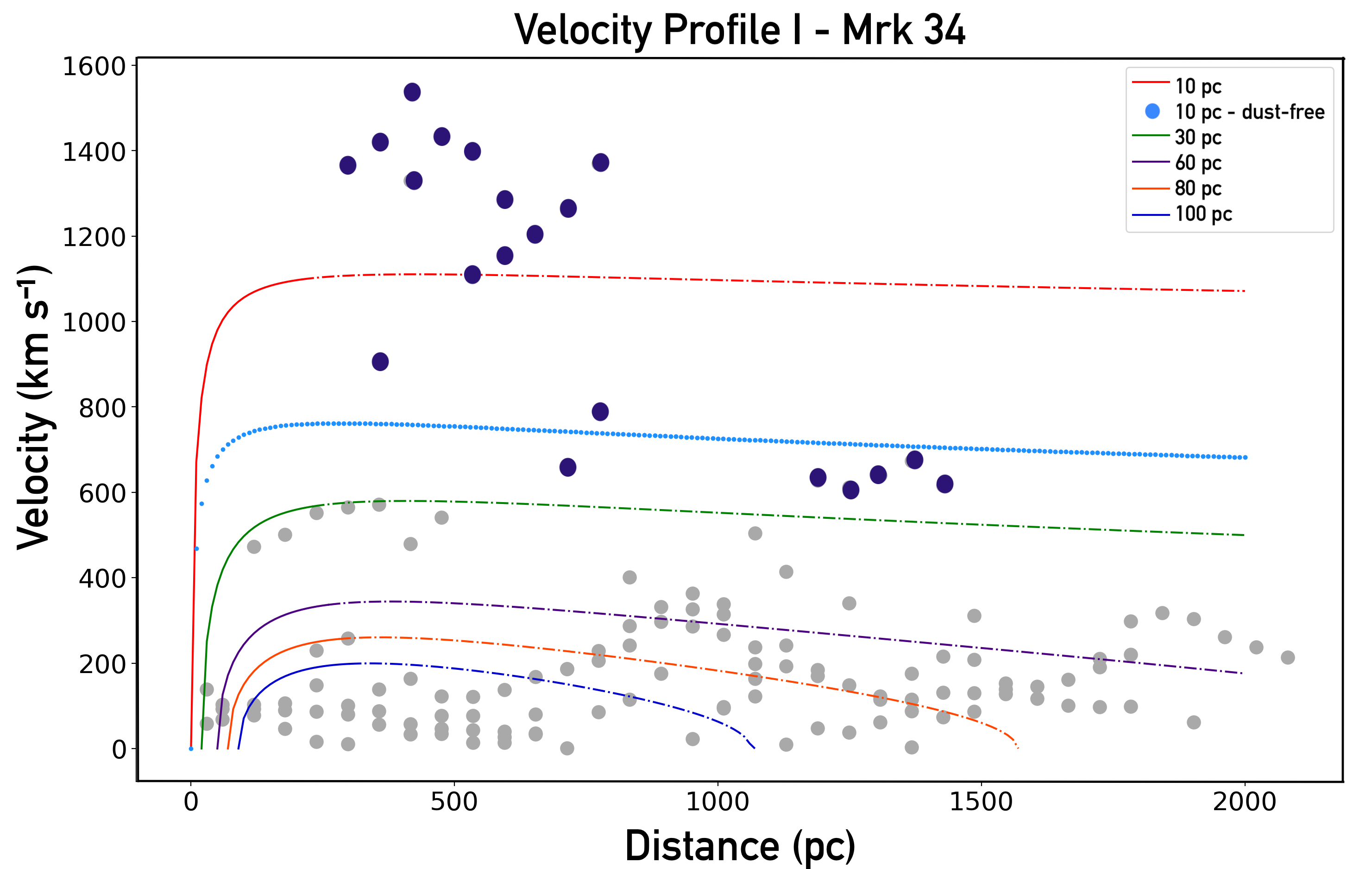}
 \end{minipage}\qquad
 \caption{Deprojected radial velocity profile for different launch radii for Mrk 477 (left panel) and for Mrk 34 (right panel). For these models we considered that the gas possesses only one force multiplier, $\mathcal{M} = 1040$, during its entire trajectory. The points represent the deprojected velocities for each target calculated using the automated emission-line fitting routine from \citet{fischer2017a}. The deep purple points represent the outflow velocities, while the gray points represent rotational velocities or are ambiguous. The solid lines represent the first 200 pc of the gas trajectory, and the dashed-dot lines represent the remainder of the gas trajectory. The values presented for the deprojected velocity were added from both sides of the nucleus. }
\label{fig:velocity_profile_1}
\end{figure*}

\begin{figure}
  \centering
   \includegraphics[width=0.5\textwidth]{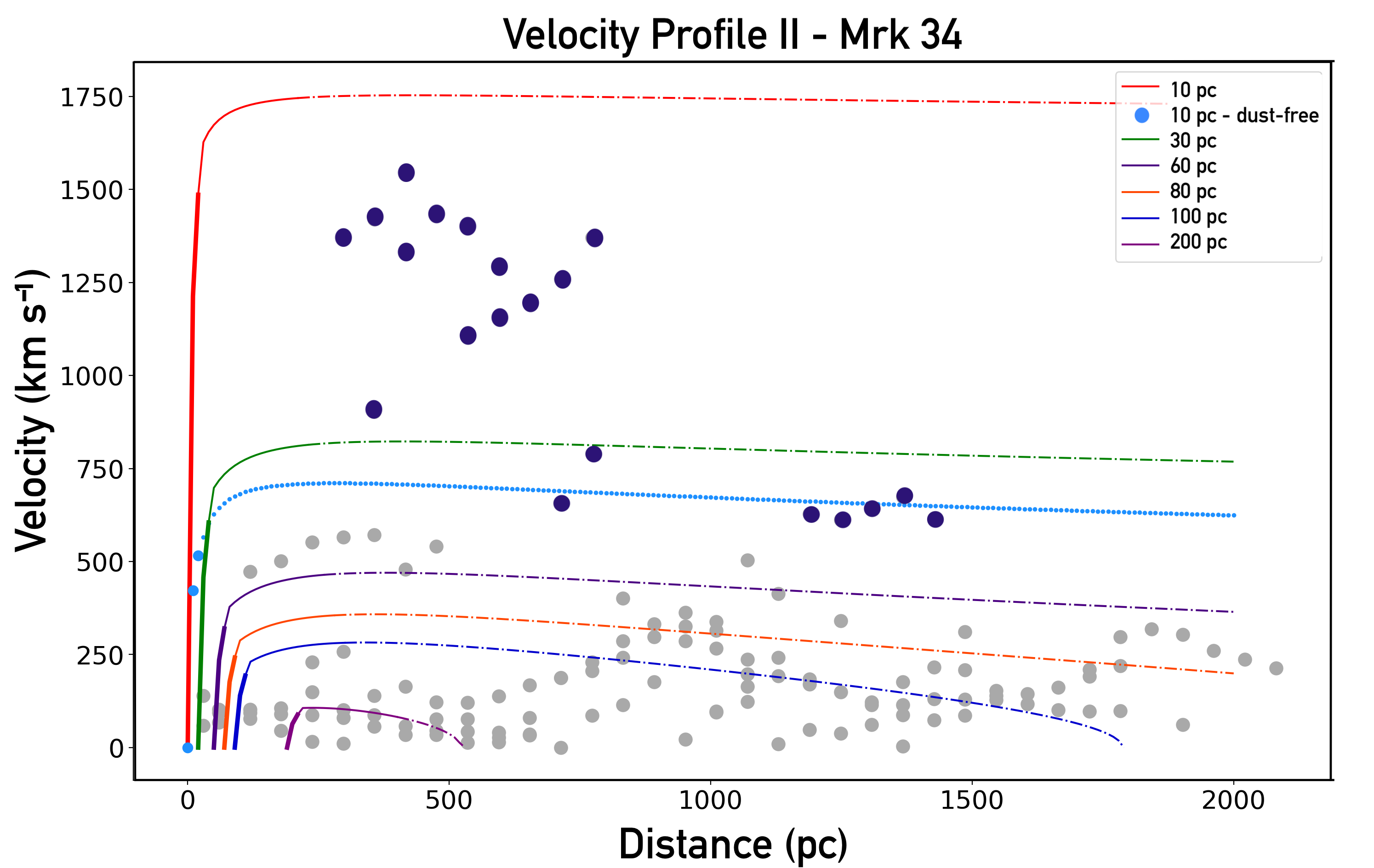}
\caption{\label{fig:velocity_profile_2} Deprojected radial velocity profile for different launch radii. For these models we consider that the gas possesses two different force multipliers, $\mathcal{M} = 4650$ and $\mathcal{M} = 1040$ , during its entire trajectory. The models are run with the assumption that the gas is in a lower ionisation state (logU=-3) for the first 20 pc (in boldface) of its journey. The points are the deprojected velocities for Mrk 34 and the curves are the velocities from our models. The solid lines represent the first 200 pc of the gas trajectory, and the dashed-dot lines represent the remainder of the gas trajectory.}
\end{figure}

\section{Mrk 477 and Mrk 34: Dynamics of the Ionised Outflows}
\label{sec:dynamics_outflows}

\subsection{Mrk 477: Radial Velocity Profiles}
\label{sec:mrk477_velocity_profile}
Using a Bayesian fitting routine (discussed in detail by \citealt{fischer2017a}) for the analysis of the STIS spectra, and the orientation of the host galaxy from the Sloan Digital Sky Survey (SDSS) images, we obtain the deprojected [O~III] velocities relative to systemic as a function of distance from the nucleus, and use this information to construct a radial velocity profile for Mrk 477 (represented by points in Figure \ref{fig:velocity_profile_1}). All of the kinematic components at each position are displayed in Figure \ref{fig:velocity_profile_1}. While most of the velocities are low, there are a small number of high velocity, i.e., $\sim$ 200 ${\rm km~s^{-1}}$, [O~III]-emitting clouds at distances $>$ 150 pc. \par 

In order to study the dynamics of this gas, we use Equation 3 to generate a grid of radial velocity profiles for different launch radii and compare these results to the deprojected velocities. For our radial velocity profiles, we determine $M(r)$ at 10 pc intervals over a range of 10 pc $\leq$ r $\leq$ 5 kpc. Our radial velocity profiles initially assume that the gas is in rotation around the gravitational center. \par 

In Figure \ref{fig:velocity_profile_1}, we present a radial velocity profile for the [O~III] gas, characterized by $\mathcal{M}$ = 1040, which is the column density weighted $\mathcal{M}$ calculated from the Cloudy models (see Equation 4), with $N_{H}=10^{21.5}$ and $logU=-2$ and an effective $L_{bol}$ of $5.91\times10^{44}$ (see section \ref{sec:force_multiplier}). We can see that the high deprojected velocities we observe in Mrk 477 (deep purple points on the left panel of Figure \ref{fig:velocity_profile_1}) can be reached if the gas is launched from close to the SMBH and there is internal dust. In this scenario, the gas would have to remain in the [O~III]-emitting state over most of its trajectory in order to achieve these high velocities. The fact that we only observe four points with high velocities, i.e., $>$  200 ${\rm km~s^{-1}}$, may be related to the evolutionary state of the AGN (see Section \ref{sec:discussion}). \citet{fischer2017a} determined that, for Mrk 573, the maximum distance that it is possible to track the [O~III] gas is $\sim$ 200 pc from the point where the molecular gas entered the bicones. Looking at the left panel of Figure \ref{fig:velocity_profile_1}, we see that it is possible that the [O~III]-emitting gas could reach the observed velocities within 200 pc from its launch point.\par 

We do not observe high velocity, i.e., centroid velocity from systemic $>$ 200 ${\rm km~s^{-1}}$, gas far from the nucleus in Mrk 477, i.e., at distances $>$ 350 pc. However, our calculations show that this AGN is capable of creating radiatively driven outflows, characterised by a single $\mathcal{M}$, provided that dust is present in the [O~III] gas (see Figure \ref{fig:velocity_profile_1}). \par

\subsection{Mrk 34: Radial Velocity Profiles}
\label{sec:mrk34_velocity_pro}
Using the same method applied to Mrk 477,  we obtain the deprojected [O~III] velocities for Mrk 34 and use this information to construct a radial velocity profile for this target. As described in \citetalias{trindadefalcao2020a}, Mrk 34 possesses the most energetic and extended outflows in our sample of study, and it is apparent that there is high velocity, i.e., $>$ 500 ${\rm km~s^{-1}}$, [O~III] gas at distances $>$ 250 pc. We generate a grid of radial velocity profiles for Mrk 34, using the method described in Section \ref{sec:mrk477_velocity_profile} and compare the results to its deprojected velocities. \par 

In Figure \ref{fig:velocity_profile_1} (right panel), we present a radial velocity profile for the [O~III] dusty gas, characterized by $\mathcal{M}$ = 1040 (see section \ref{sec:force_multiplier}) and an effective $L_{bol}$ of $5.35\times10^{44}$ ${\rm ergs~s^{-1}}$ (see section \ref{sec:force_multiplier}). If the gas remains in the [O~III] state, i.e., if it possesses a constant $U$, $N_{H}$ has to decrease with distance, as follows:

\begin{equation}
    \frac{N_{H_{1}}}{N_{H_{2}}} = (\frac{r_{1}}{r_{2}})^{-\frac{4}{3}}
\end{equation}

\noindent where $N_{H_{i}}$ is the column density of the i-th cloud of gas and $r_{i}$ is its radius. However, as shown in Figures \ref{fig:force_multiplier} and \ref{fig:bolometric_luminosity}, $\mathcal{M}$ is fairly constant and the effective $L_{bol}$ actually increases at lower $N_{H}$, therefore, we can approximate what the trajectories are by assuming a constant $N_{H}$.  \par 

We can see that, if the [O~III] gas is launched from distances close to the SMBH, for example, $\leq$ 70 pc, it can reach the high deprojected velocities we observe, at distances $<$ $\sim$ 800 pc. However, as in Mrk 477, the gas would have to remain in the [O~III]-emitting state over most of its trajectory in order to achieve these high velocities. In addition, according to \citet{crenshaw2015a} and \citet{fischer2017a}, the amount of gas that we detect in outflow could not be explained if the gas was launched from such small radial distances. Specifically for Mrk 34, \citet{revalski2018a} show that the spatially-resolved mass outflow rate exceeds the nuclear rate. Therefore, even though our calculated trajectories show that the [O~III] gas could reach the observed velocities, the scenario presented in Figure \ref{fig:velocity_profile_1} (right panel) may not be physically probable.\par

Another possibility is that the gas does not start its trajectory as [O~III]-emitting gas. \citet{fischer2017a} suggested that cold gas could rotate into the solid angle illuminated by the AGN, ionising quickly while being accelerated, which is consistent with in-situ acceleration. In this case, the gas may start off in a lower ionisation, higher density state than the [O~III]-emitting gas and, then, evolve into [O~III] gas while the acceleration rate drops (e.g., \citealt{hagino2015a,nomura2016a}, which discuss radiative acceleration of UFOs). For example, if the gas in this initial stage is characterized by a log$U$= -3, it would have a higher opacity, corresponding to
$\mathcal{M}$ $\sim$ 4700 (see Figure \ref{fig:force_multiplier}). If the gas begins to be accelerated while in this lower ionisation/denser state, it might be at a high enough
velocity when it expands to the point where it is [O~III] gas, which is what we detect in outflows.\par

However, as shown in Figure \ref{fig:force_multiplier}, a cloud of gas characterized by a ionisation parameter of log$U$= -3  would have a lower column density than $10^{21.5}$. If these are simply dense knots clouds of gas, mass conservation would suggest that they would not evolve into the high column density [O~III] gas we detect. Therefore, the scenario of a low ionisation/high density cloud expanding into a [O~III] cloud with log$N_{H}$ = 21.5 is not physically consistent. \par

On the other hand, if neutral gas is continuously flowing into the NLR (see \citealt{fischer2017a}), outer parts could be ionised, with the gas initially in a low-ionisation state. As this gas expands it forms the [O~III] wind. Thus, the inflow provides a reservoir of material which evolves into [O~III] emitting gas. To illustrate this scenario, we calculate the trajectory of gas which is initially characterised by log$U$= -3 and expands until it reaches a log$U$= -2 state (see Figure \ref{fig:velocity_profile_2}). We can see that the gas can achieve the observed high velocities we detect in Mrk 34 if it is launched from very small distances from the SMBH, i.e., $<$ 20 pc. It is possible that the gas could have started in a lower ionisation state in Mrk 477 as well, however we do not observe any obvious effect on the radial velocity profiles to support this hypothesis.\par 

 On the other hand, looking at Figure \ref{fig:velocity_profile_1} (right panel) and \ref{fig:velocity_profile_2}, we notice that there can be high velocity gas at distances $>$ 200 pc from the point of origin for several launch radii. For example, we detect emission at $\sim$ 750 pc with velocities of $\sim$ 500 ${\rm km~s^{-1}}$, that have a launch radius of $\sim$ 200 pc. This means that either the [O~III] gas would have to remain in the [O~III] state for much further than in Mrk 573 or there is another mechanism which is responsible for high velocity [O~III] gas we detect at large distances. We  explore this possible mechanism in Section \ref{sec:entrainment}.
 
 \subsection{Expansion Rates and Confinement}
\label{sec:expansion_rates}

Based on the radial velocity profile for Mrk 34 presented in the right panel of Figure \ref{fig:velocity_profile_1} and Figure \ref{fig:velocity_profile_2}, we know that if the gas has a $\mathcal{M}$ of 1040 it will not reach the observed velocities we see at $\sim$ 1.2 kpc, unless it is launched from close to the AGN (as shown in Figure \ref{fig:velocity_profile_1}). However, as the distance to the SMBH increases, the density drops and the size of the cloud expands. If the [O~III] gas is not confined, it would expand rapidly into X-ray gas. \par

We can calculate the time it takes for the [O~III]-emitting gas to expand, i.e., the free-expansion times. To do so, we assume that the motion of the outer edge of the cloud is given as follows \citep[e.g.,][]{osterbrock2006a}:
\begin{equation}
 r_{e} = r_{o} + u_{e}t   
\end{equation}
\noindent where $r_{o}$ is the initial radius of the cloud, and $u_{e}$ is the velocity of the edge of the gas cloud. In this case, $u_{e}$ is equal to the sound speed:
\begin{equation}
    u_{e} = \sqrt{\frac{\gamma~k~T_{o}}{\mu_{o}~m_{H}}}
\end{equation}
\noindent where $\gamma = 1$, for isothermal expansion, k is the Boltzmann constant, $T_{o}$ is the temperature calculated by the Cloudy model for the lower ionisation state, in this case $\sim$ $10^{4}$~K, $\mu _{o} \sim 0.6$ is the mean mass per particle and $m_{H}$ is the mass of hydrogen.\par

When the gas expands and becomes more ionised, its density decreases as $1/U$. If we assume that the cloud expands equally in all directions, for a factor of 10 increase in volume, corresponding to a change from log$U$= -2 to log$U$= -1, we have: 
\begin{equation}
    r_{e} = 2.15\times r_{o}
\end{equation}
\noindent The value of $r_{o}$ can be calculated for each position, which gives us $r_{e}$. Using Equation 7, we calculate the time it takes for the lower ionisation gas, i.e., log$U$= -2, to expand to become a higher ionised gas, i.e., log$U$= -1.\par 

For a cloud of gas that is launched from $r$ = 50 pc, the free-expansion time from a log$U$ = -2 state to a log$U$ = -1 state is \textbf{$2.5\times10^{10}$}~s. For example, for a $v = 1000 {\rm km~s^{-1}}$ this would mean that the cloud would be able to move less than 1 pc during this time. As noted before, \citet{fischer2017a} found that the [O~III] gas can travel $\sim$ 200 pc from its origin point, which suggests that the gas is partially confined, due to radiation pressure and/or ram pressure, but it is plausible that it will eventually expand into X-ray gas. \par

\section{The origin of X-ray winds}
\label{sec:origin_Xrays}
As mentioned above for our sample of QSO2s, \citet{fischer2018a} found that the influence of the AGN changes with increasing distance from the SMBH. In the most distant regions the gas presents low central velocities, which is consistent with rotation, but high FWHM, which they identified as "disturbed" or turbulent gas. We suggest that this phenomenon may be due to the impact of X-ray winds.\par

\subsection{Kinetic Energy Density Analysis}
\label{sec:kinetic_energy}
If the disturbance of the [O~III] gas \citep{fischer2019a} in Mrk 34 is due to interaction with an X-ray wind, we can assume that the wind is depositing kinetic energy into the [O~III] gas, which suggests that X-ray wind and the [O~III] gas may possess similar kinetic energy densities. Here we are assuming an "ideal case" scenario, in which we ignore any sort of energy dissipation (e.g., \citealt{bottorff2002a} and references therein). If this is the case, we can estimate the properties of an X-ray wind that might create this disturbance. \par

We calculate kinetic energy density of the disturbed gas, $u_{_{[O~III]}}$, as follows:
\begin{equation}
    u_{_{[O~III]}} \sim \frac{1}{2} n_{H}m_{p} \sigma^2
\end{equation}
\noindent where $n_{H}$ is the density of the [O~III] gas from the photoionisation models and $\sigma$ is the velocity dispersion of the gas (see \citetalias{trindadefalcao2020a}), which ranges from $3.4\times10^{1}~{\rm km~s^{-1}}$ to $6.6\times 10^{2}~{\rm km~s^{-1}}$.\par

\begin{figure*}
  \centering
 \begin{minipage}[b]{0.45\columnwidth}
  \advance\leftskip-5cm
  \includegraphics[width=9cm, height=6cm]{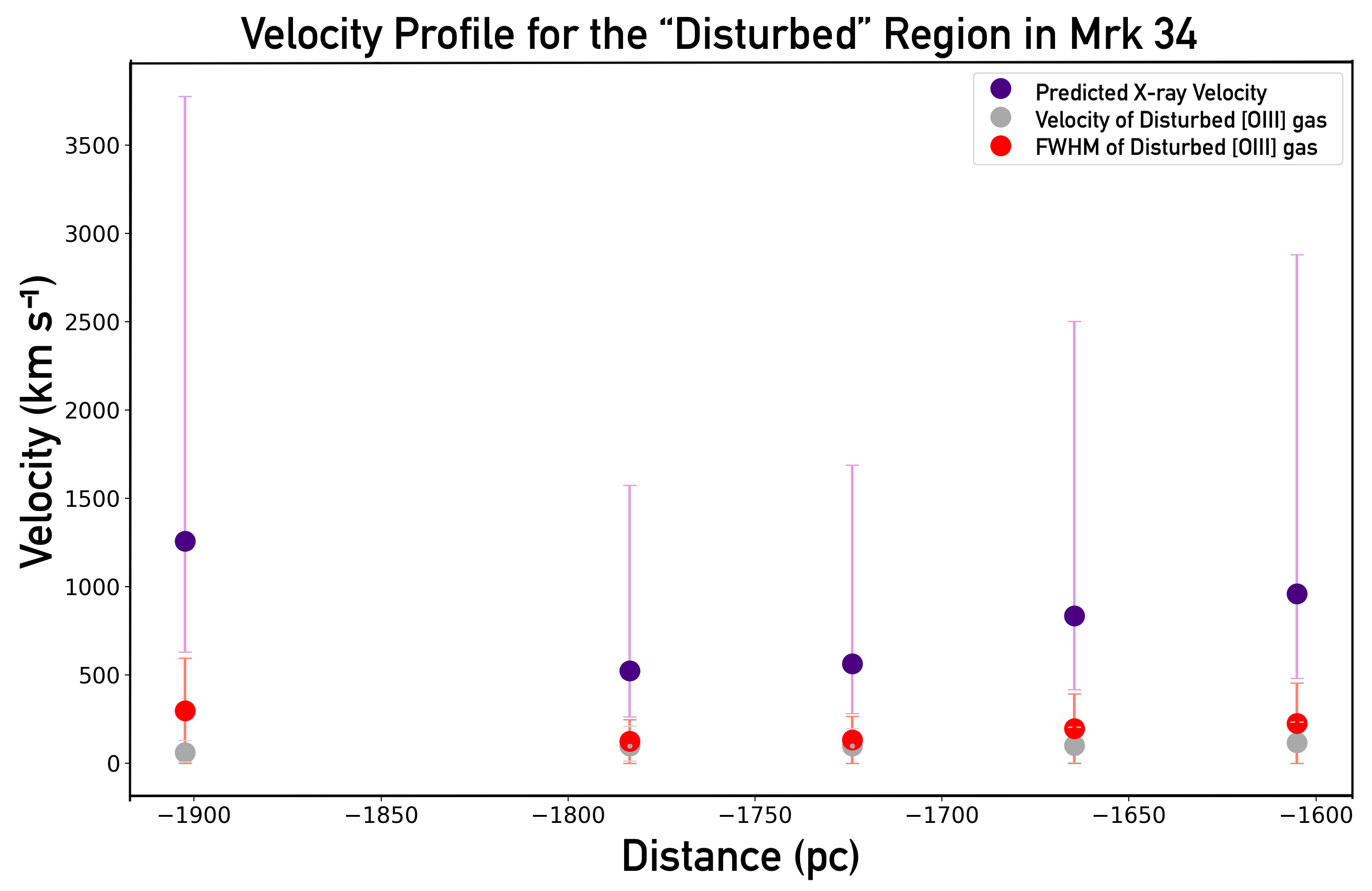}
 \end{minipage}\qquad
 \begin{minipage}[b]{0.45\columnwidth}
  \includegraphics[width=9.2cm, height=6.05cm]{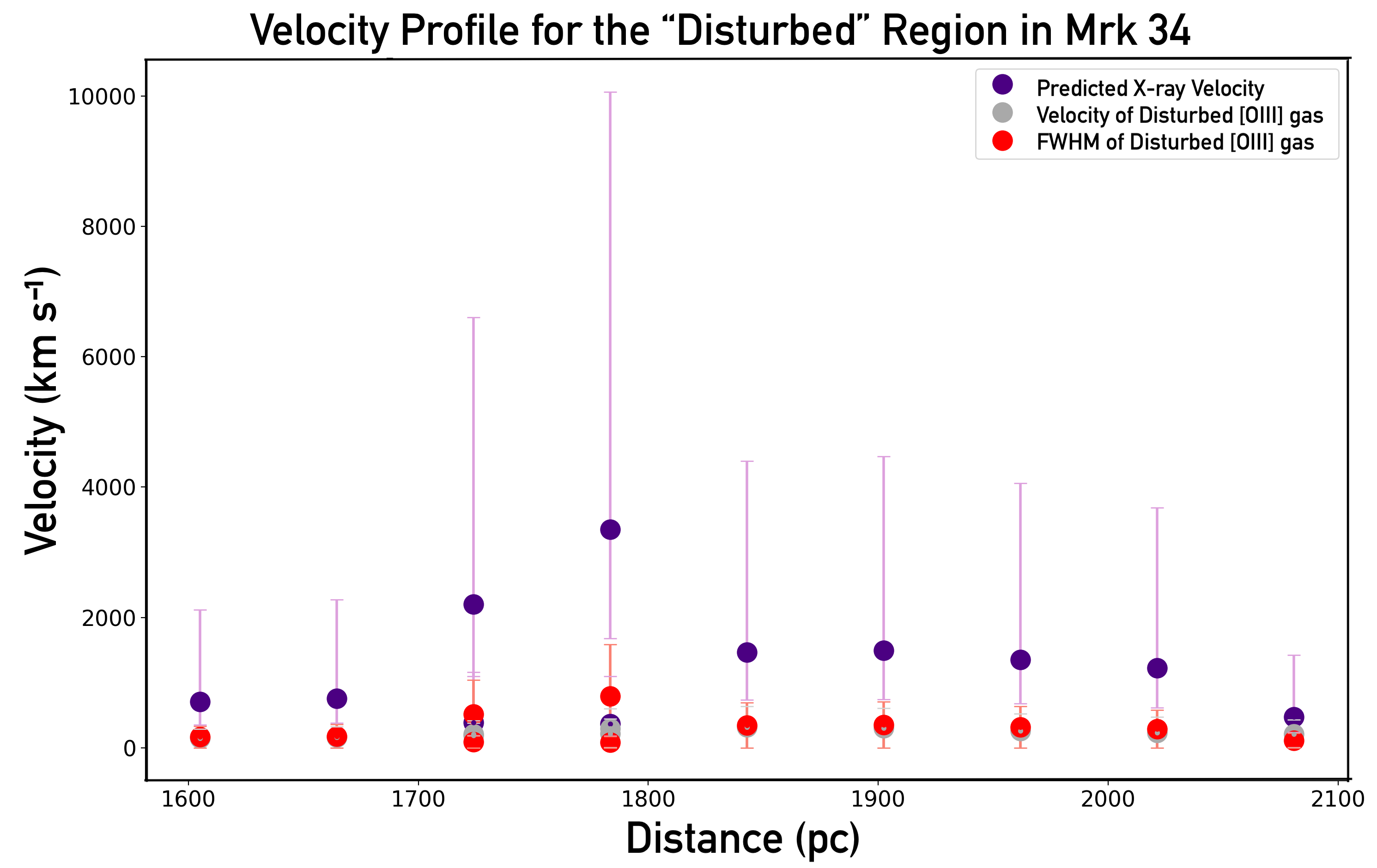}
 \end{minipage}\qquad
 \caption{Predicted radial velocity profile for the X-ray winds (in purple). The gray points represent the velocity centroid of the disturbed [O~III] gas detected in Mrk 34 \citep{fischer2018a}, and the red points represent the FWHM of the [O~III] disturbed gas. The uncertainties for the predicted X-ray velocities are calculated following the method described in \citetalias{trindadefalcao2020a}, i.e., a factor of 2 in our assumed ionisation parameter. The uncertainties for the velocity of the disturbed [O~III] gas and its FWHM are calculated using the results of the Bayesian fitting routine, as discussed by \citealt{fischer2017a}. The left panel represents the points to the southwest and the right panel represent the points to the northwest.} 
\label{fig:Xray_o3_velocity}
\end{figure*}

For the X-ray wind, we assume log$U$ = 0.0 \citep{turner2003a,armentrout2007a,kraemer2015a, kraemer2020a}, which is  consistent with models of the NLR X-ray emission-line gas. Using the fact that
 \begin{equation}
    n_{x} = \frac{U_{_{[O~III]}}}{U_{x}} n_{H}
 \end{equation}
 \noindent where $U_{_{[O~III]}}$ is the ionisation parameter of the [O~III]-emitting gas and $U_{x}$ is the ionisation parameter of the X-ray wind, we are able to estimate the densities for the X-ray gas, $n_{x}$. Assuming that the kinetic energy density of the [O~III] gas is equal to the kinetic energy density of the X-ray gas, then:  
 \begin{equation}
     v_{x} = \sqrt{\frac{2u_{_{[O~III]}}}{n_{x} m_{p}}}
 \end{equation}
\noindent gives us the velocity of the X-ray wind. We present our results for the radial velocity profile of the X-ray winds and the [O~III] disturbed gas in Mrk 34 in Figure \ref{fig:Xray_o3_velocity}.\par

Using Equation 12, we compute the velocities for the X-ray gas in Mrk 34. These predicted velocities are very high, ranging from $3.7\times 10^{2}~{\rm km~s^{-1}}$ to $3.3\times10^{3}~{\rm km~s^{-1}}$ (see Figure \ref{fig:Xray_o3_velocity}). Considering the low force multiplier values for highly ionised gas (see Figure \ref{fig:force_multiplier}), the X-ray gas could not be radiatively accelerated to these velocities, even if it originated closer to the AGN (see Figure \ref{fig:velocity_profile_3} and discussion in Section \ref{sec:entrainment}). A more plausible scenario is that the wind begins as [O~III]-emitting gas launched close to the SMBH, e.g., $<$ 100 pc, and subsequently expands over its trajectory, eventually evolving into an X-ray wind. For Mrk 34, considering that the size of the outflow region is $\sim$ 1.6 kpc, and assuming that the X-ray winds were launched from 10 pc and possess a velocity of $\sim$ 960 ${\rm km~s^{-1}}$ at this position (see Figure \ref{fig:velocity_profile_2}), our calculations show that this target would have entered a high state of activity in which it is capable of launching outflows $\sim$ 1 million years ago.\par

\subsection{Entrainment}
\label{sec:entrainment}

\begin{figure}
  \centering
 \includegraphics[width=0.5\textwidth]{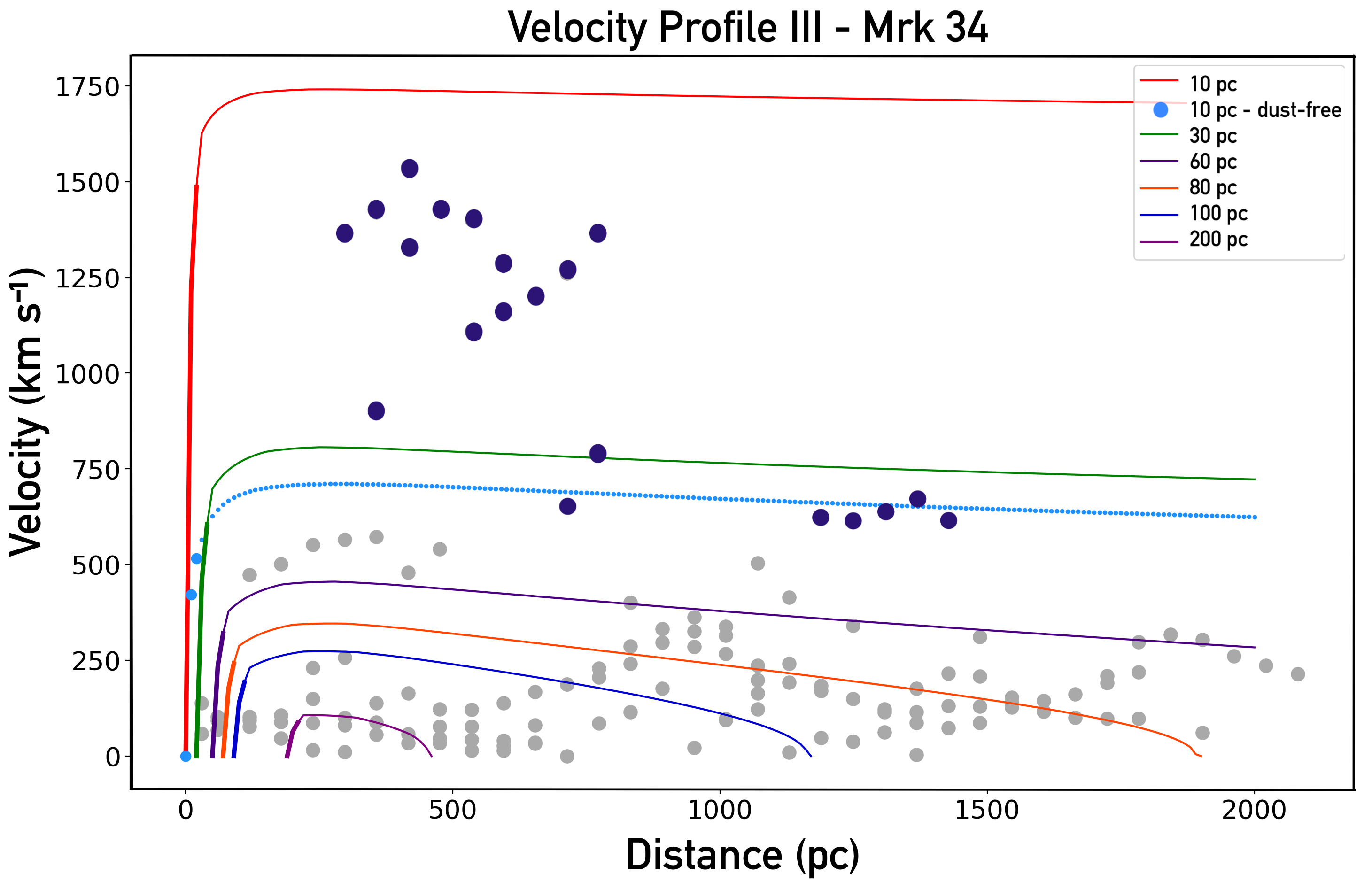}
\caption{\label{fig:velocity_profile_3} Deprojected radial velocity profiles for different launch radii. For these models we considered that the gas possesses five different force multipliers over its entire trajectory, i.e., $\mathcal{M} = 4650$, $\mathcal{M} = 1040$, $\mathcal{M} = 192$, $\mathcal{M} = 112$ and $\mathcal{M} = 83$. We assume that the gas is in a lower ionisation state (log$U$= -3) for the first 20 pc of its journey (in  boldface), and that it remains in the other ionisation states (log$U$= -2, log$U$= -1 and log$U$= -0.5) each for 100pc, and in the highest ionisation state (log$U$= 0.0) for the rest of its trajectory. The changes in $\mathcal{M}$ result from the decrease in density of the expanding clouds (see section \ref{sec:force_multiplier}). The points are the deprojected velocities for Mrk 34 and the curves are the velocities from our models.}
\end{figure}

In addition to the disturbed kinematics, other possible evidence for an X-ray wind in Mrk 34 is the presence of high velocity [O~III] gas at $\sim$ 1.2 kpc, as shown in Figures \ref{fig:velocity_profile_1} (right panel), \ref{fig:velocity_profile_2} and \ref{fig:velocity_profile_3}. Although it is possible that this gas originates close to the AGN and remains in the [O~III]-emitting state over its whole trajectory, this scenario is inconsistent with in situ acceleration and with the conclusions of \citet{fischer2017a} for Mrk 573 (see discussion in Section \ref{sec:mrk477_velocity_profile}). A more plausible scenario is that it was launched relatively close to where it is detected. This leaves us with the question: how is this [O~III] gas accelerated to high velocities at these large distances? \par

One possibility is that the [O~III] gas is accelerated in-situ via entrainment by X-ray winds which originate closer to the AGN \citep[e.g.,][]{shimizu2019a}. Multi-wavelength observations of nearby galaxies show that galactic winds are multiphase, with cold ($\sim$ $10^{4}~K$) gas co-existing with hot($\sim$ $10^{6}~K$) gas, and moving outward at very high speeds. For example, in the Milky Way a large population of fast-moving, cold gas clouds have been detected \citep{wakker1997a}; these seem to be disturbed throughout the Galactic halo (e.g., \citealt{diteodoro2018a}, and references therein). Absorption line studies at higher redshifts show similar results (e.g., \citealt{rudie2019a}). One important question is whether a hot wind can accelerate dense clouds of gas to the observed velocities via hydrodynamic ram pressure (e.g., \citealt{gronke2019a}, and references therein). When it comes to this type of acceleration, the main problem is to compare the acceleration times of the clouds to their destruction times. This can be demonstrated by a timescale argument. In our study, we estimate the acceleration timescales derived from our observations and simple analytic calculations. \par 

If an X-ray wind is launched at 10 pc as low-ionisation gas, it would reach velocities at $\sim$ 1.2 kpc that are much greater than those of the [O~III] gas (see Figure \ref{fig:velocity_profile_3}). In fact, this process is necessary in order to get the [O~III] gas accelerated to the observed velocities by entrainment.\par 

\begin{figure}
  \centering
  \includegraphics[width=0.5\textwidth, height=5.5cm]{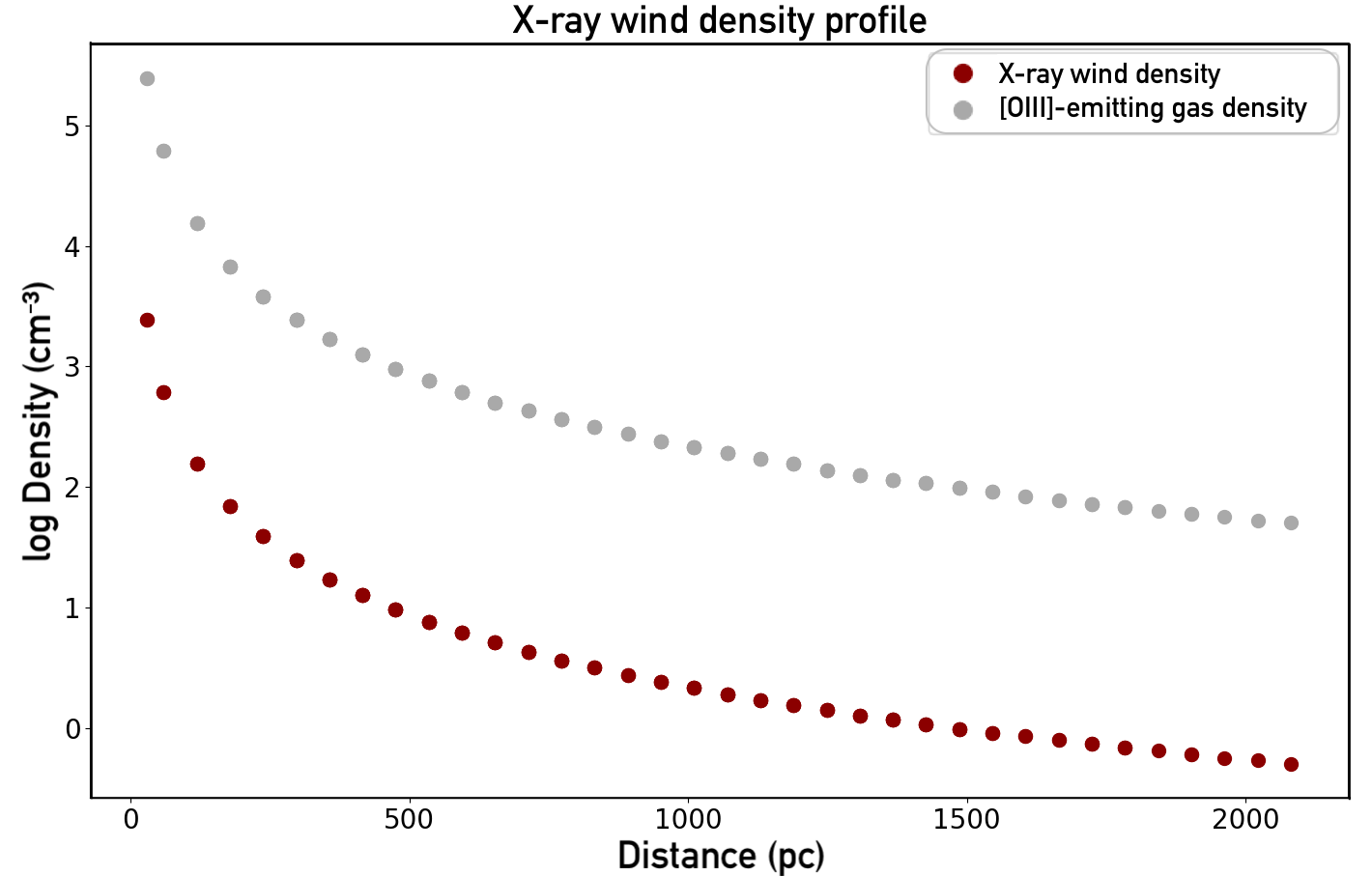}
\caption{\label{fig:Xray_density} Radial density profile for the X-ray wind (in red). The densities were calculated using equation 10, with an ionisation parameter of log$U$= 0.0. The gray points represent the [O~III]-emitting gas density profile computed in \citetalias{trindadefalcao2020a}, and are shown for comparison. }
\end{figure}

To estimate the distance over which acceleration by entrainment can occur, we construct a radial density profile for the X-ray wind in Mrk 34, as shown in Figure \ref{fig:Xray_density}. For example, assuming log$U$ = 0.0 \citep{turner2003a,armentrout2007a,kraemer2015a, kraemer2020a}, $n_{x} = 1.4~{\rm cm^{-3}}$ at $\sim$ 1.2 kpc. \par 

As shown by \citet{everett2007a}, the drag force for clouds injected into winds and accelerated by ram pressure is:
\begin{equation}
    F_{drag_{[O~III]}} = \rho_{x}(v_{x}- v_{_{[O~III]}})^2 A_{_{[O~III]}}
\end{equation}
where $\rho_{x}$ is the mass density of the wind, $v_{_{[O~III]}}$ is the initial velocity of the [O~III] clouds of gas, and $A_{_{[O~III]}}$ represents the cross sectional area of the cloud.\par 

The acceleration for a cloud of [O~III] gas is:
\begin{equation}
 \begin{split}
    a_{drag_{[O~III]}} = \frac{\rho_{x}}{\rho_{_{[O~III]}}}~\frac{\frac{3}{4}(v_{x} - v_{_{[O~III]}})^2}{R_{_{[O~III]}}} =\\ \frac{n_{x}}{N_{{H}_{[O~III]}}}~ \frac{3}{4}~(v_{x} - v_{_{[O~III]}})^2
 \end{split}
\end{equation}
\noindent where $R_{_{[O~III]}}$ is the radius of an individual cloud. We assume a column density of $N_{{H}_{[O~III]}} = 10^{21.5}~{\rm cm^{-2}}$ for our models (see \citetalias{trindadefalcao2020a}).\par 

We calculate the velocity of the [O~III] gas as a function of distance as follows:
\begin{equation}
 \begin{split}
   \frac{dv(r)}{dr} = 3420L_{44}\frac{\mathcal{M}}{v(r)r^{2}} - 4.3\times10^{-3}\frac{M(r)}{v(r)r^{2}} \\
   +2.3\times10^{-2}\frac{n_{x}}{N_{H}} \frac{(v_{x}-v_{_{[O~III]}})^{2}}{v_{_{[O~III]}}}
 \end{split}
\end{equation}
In order to calculate $v(r)$ in Equation 15, we employ the 4th order Runge-Kutta method to solve the differential equation. \par 

\begin{figure}
  \centering
  \includegraphics[width=0.5\textwidth, height=5.8cm]{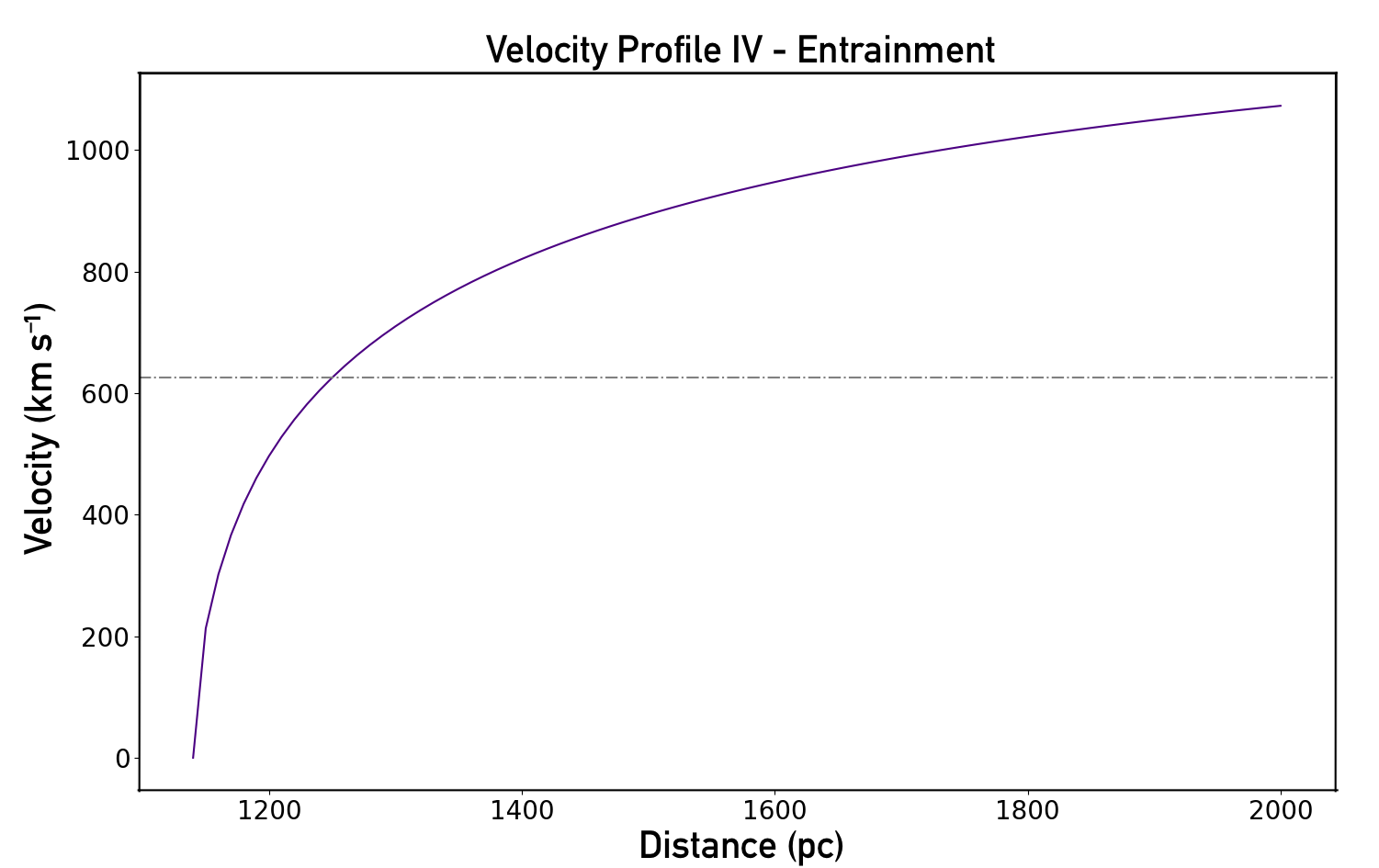}
\caption{\label{fig:velocity_profile_4} Radial velocity profile for entrainment in Mrk 34. For this model, the [O~III]-emitting gas has an initial velocity $v_{_{[O~III]}}$ = 53 ${\rm km~s^{-1}}$ at 1.1 kpc, calculated as described in section \ref{sec:mrk477_velocity_profile}, and is dragged by the X-ray wind for $\sim$ 100 pc, until it reaches the observed velocities, i.e., $\sim$ 615 ${\rm km~s^{-1}}$. The gray dashed line represents the final velocity of the [O~III]-emitting gas at $\sim$ 1.2 kpc. }
\end{figure}

As shown in Figure \ref{fig:velocity_profile_4}, in order to achieve the velocities observed at $\sim$ 1.2 kpc, the [O~III] gas would have to have been carried by an X-ray wind with a velocity of $v_{x} = 1.5\times10^{3}~{\rm km~s^{-1}}$ for 100 pc. As noted above, the X-ray wind would have to have been launched $\sim$ 10 pc from the SMBH to possess this velocity (see Figure \ref{fig:velocity_profile_3}).\par

As Equation 14 shows, $a_{drag_{[O~III]}}$ is inversely proportional to $N_{H}$. This indicates that optically thin [O~III] filaments, i.e., clouds with lower column densities, can be entrained more efficiently. This suggests that the higher velocity gas we observe in Mrk 34 could be matter-bounded. Note that in matter-bounded gas line ratios such as [Ne~V]~3426\AA/H$\beta$ and He~II~4686\AA/H$\beta$ would increase compared to those in radiation-bounded gas \citep{binette1996a, kraemer1998a}.\par 

We also check the stability of the clouds by comparing their Kelvin-Helmholtz timescale and their acceleration time. The Kelvin-Helmholtz timescale is given by \citep{everett2007a}:

\begin{equation}
t_{_{KH}} \sim \frac{R_{_{[O~III]}}}{v_{_{[O~III]}}} (\frac{n_{_{[O~III]}}}{n_{x}})^{1/2}
\end{equation}

\noindent where $R_{_{[O~III]}}$ is the radius of the [O~III] cloud. \par 
The acceleration time is given by:

\begin{equation}
t_{acc} \sim \frac{d}{v}
\end{equation}

For the [O~III] clouds at $\sim$ 1.2 kpc in Mrk 34, the Kelvin-Helmholtz timescale is $1.9\times10^{12}$~s, and the acceleration time is $6.6\times10^{12}$~s. This means that the clouds must be resistant to either Kelvin-Helmholtz or Rayleigh-Taylor instabilities to reach the observed velocities. For example, it has been proposed that magnetic field wrapping around the cloud discontinues Kelvin-Helmholtz instabilities
via magnetic tension \citep[e.g.,][]{mccourt2015a, zhang2017a}.\par 

The clouds "crushing time", i.e., the destruction timescale of the [O~III] clouds of gas which are impinged by an X-ray wind \citep{gronke2019a}, is:

\begin{equation}
t_{cc} \sim \frac{R_{_{[O~III]}}}{v_{x}} (\frac{n_{_{[O~III]}}}{n_{x}})^{1/2}
\end{equation}
For clouds at $\sim$ 1.2 kpc, $t_{cc}$ = $7.3\times10^{11}$~s, i.e., $t_{cc}$ $<$ $t_{acc}$, which means that the clouds of [O~III]-emitting gas would be destroyed before being accelerated. However, several solutions for this impasse have been suggested \citep{gronke2019a}. \par 

It has been proposed that the cooling plays an important role in stabilizing the clouds of gas, extending their lifetime \citep{klein1994a, scannapieco2015a, schneider2017a}. \citet{begelman1991a} and \citet{gronke2018a} suggest that if the cooling time of the mixed gas, i.e., gas composed of the [O~III] clouds and the X-ray wind, is less than the crushing-time then the mixing layer will cool sufficiently fast and the lifetime of the clouds of gas will be extended. Another possibility is that the energy associated with the instabilities can be radiated away when the mixture zone reaches very high temperatures. \par

\section{X-ray wind analysis}
\label{sec:Xray_wind}
\subsection{X-ray Radial Flux Distribution}
\label{sec:Xray_flux}
In sections \ref{sec:kinetic_energy} and \ref{sec:entrainment} we present the idea that X-ray winds have a significant effect on the AGN outflows we detect in Mrk 34. In section \ref{sec:kinetic_energy} we suggest that the disturbed kinematics that \citet{fischer2018a} observe at large distances from the SMBH for this target is due to the effect of the X-ray wind on the [O~III] gas. This is supported by the fact that the \textit{Chandra}/ACIS image shows extended soft X-ray emission, which suggests the presence of highly ionised gas. In addition, in section \ref{sec:entrainment}, we suggest that X-ray winds could be entraining gas, resulting in [O~III] outflows far from where acceleration by radiation pressure could be effective. \par

In the entrainment scenario the X-ray winds hit small clouds or filaments of [O~III] gas, effectively "sweeping" them up. On the other hand, in the disturbance scenario the X-ray winds may be hitting a dense spiral of gas resulting in compression rather than acceleration, as discussed by \citet{fischer2019a}.\par 

If the X-ray wind has a significant effect on the NLR dynamics, it suggests that there is large mass of X-ray emitting gas. In order to study this scenario, we use Cloudy models to predict the X-ray emission lines and the spatial distribution of the gas, and compare to the observed spatial distribution derived from \textit{Chandra} image.\par

To estimate the amount of X-ray gas in Mrk 34, we base our analysis on \citet{kraemer2020a}, who found that the radial mass profile for the X-ray gas is roughly the same as that of the [O~III] gas for the Seyfert galaxy NGC 4151. \par 

\begin{figure*}
  \centering
 \begin{minipage}[b]{0.45\columnwidth}
  \advance\leftskip-5cm
  \includegraphics[width=9cm, height=6cm]{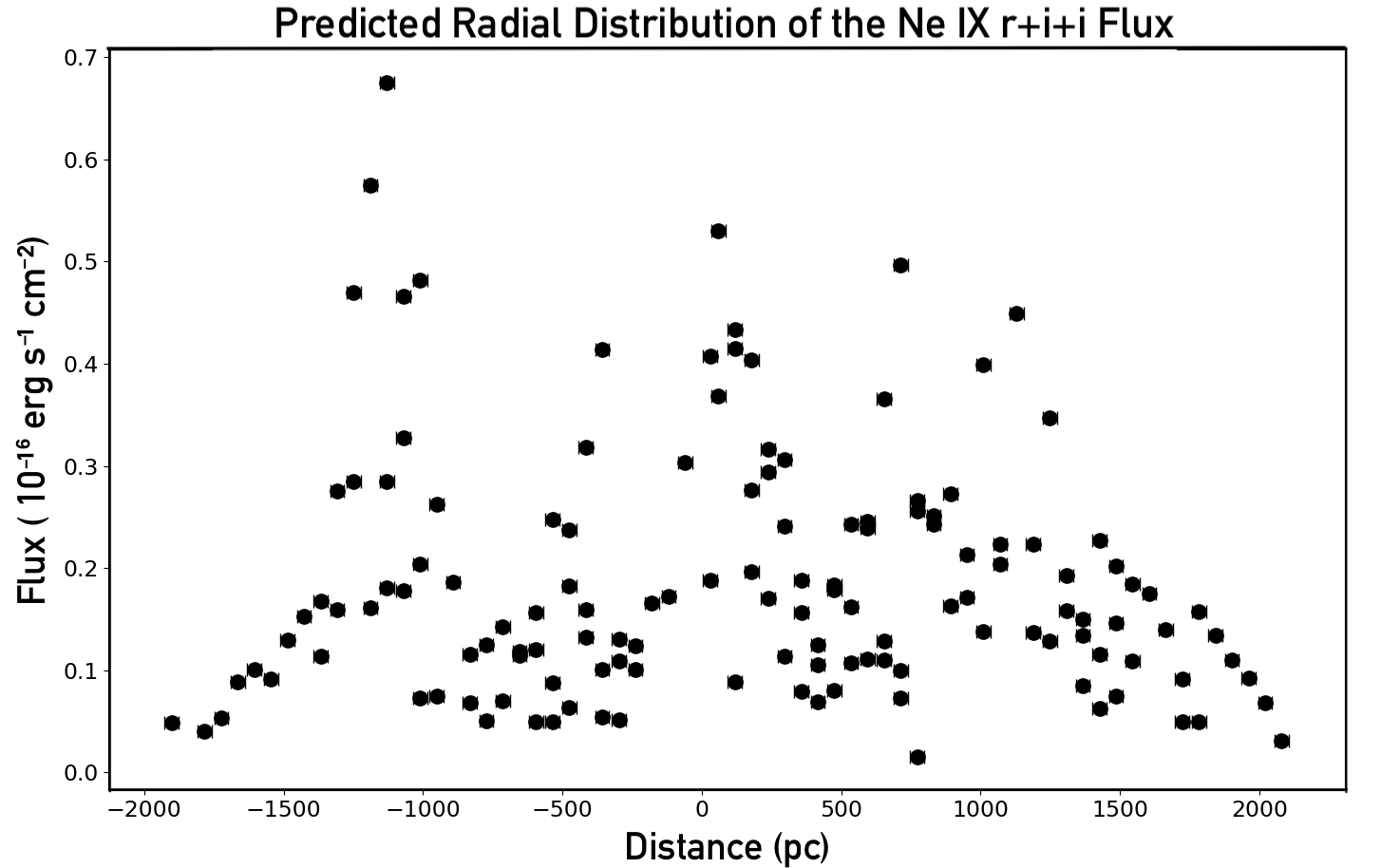}
 \end{minipage}\qquad
 \begin{minipage}[b]{0.45\columnwidth}
  \includegraphics[width=9cm, height=6cm]{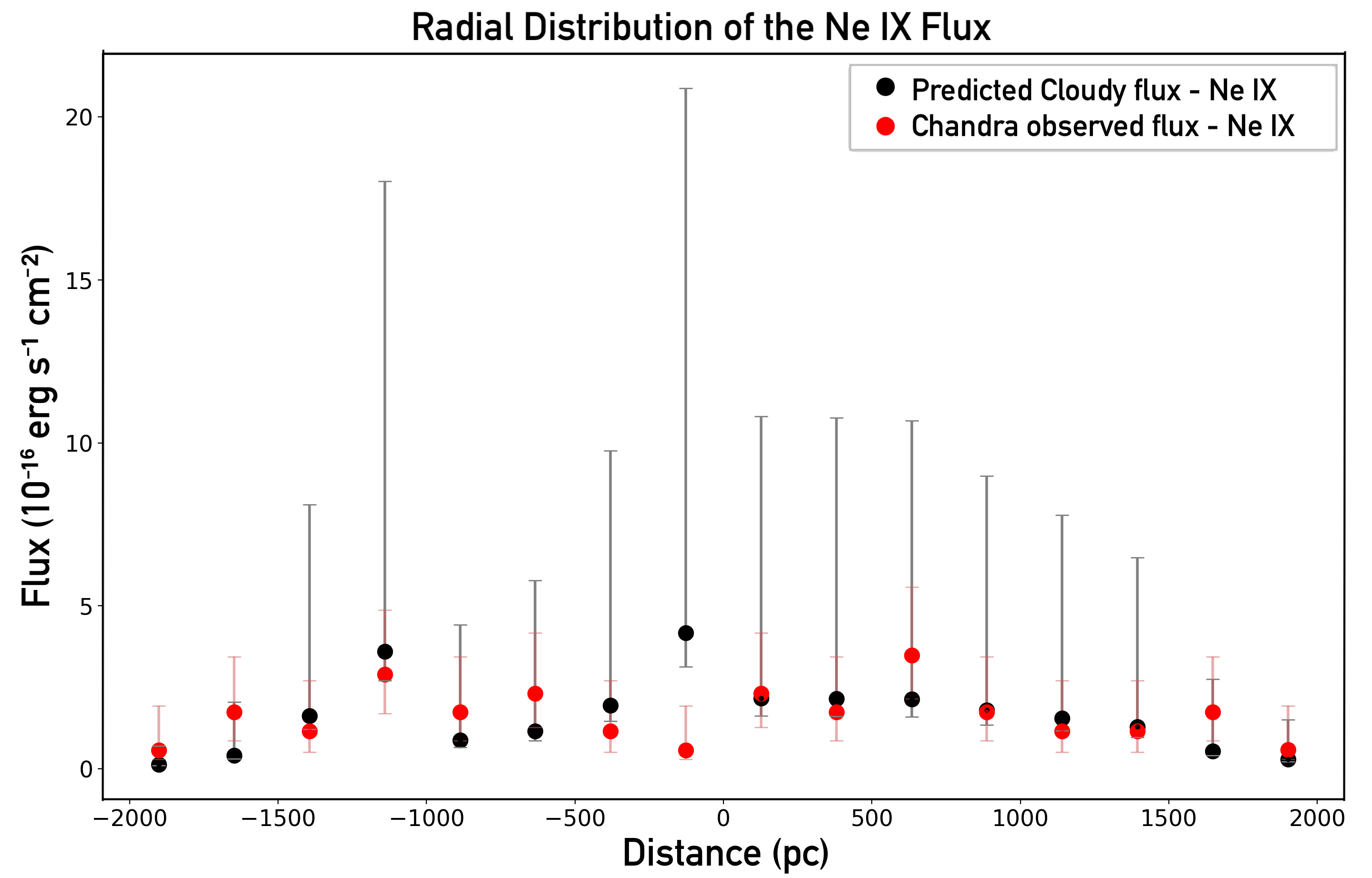}
 \end{minipage}\qquad
 \caption{Left panel: Computed radial distribution of the Ne~IX flux for Mrk 34 . The fluxes are obtained using the same analysis described in Paper I, section 3, and we assume that the mass of [O~III]-emitting gas and Ne~IX-emitting gas are roughly the same. The negative values are to the southeast and the positive values are to the northwest. For these models, we assume an X-ray component where O~VII gas peaks, i.e., a log$U$= 0.0 model. Right panel: Comparison between the predicted Cloudy spatially-integrated flux for the Ne~IX triplet and the observed flux derived from the \textit{Chandra}/ACIS image for Mrk 34. The y-axis error bars in our models were derived as described in \citetalias{trindadefalcao2020a}, including the surface areas and radial distances. The x-axis error bars for the \textit{Chandra} data points and for the points in our model are +/- 0.36 kpc, and we binned our model points to match the ACIS resolution. To obtain better X-ray spatial resolution would require the use of the Advanced X-ray Imaging Satellite (AXIS) or, preferably, the Lynx X-ray Observatory.}
\label{fig:NeIX_flux}
\end{figure*}

If we assume that Mrk 34 has the same relation between [O~III] and X-ray gas as in NGC 4151, it is possible to predict the flux of X-ray emission lines, such as those of O~VII and Ne~IX. We use the Cloudy models to get the line fluxes. In doing so, we constrain the column density, $N_{H}$, by requiring a fixed extraction bin of 50 pc in the radial direction, which means that $N_{H}$ decreases with the distance from the SMBH, as the density decreases \citep{kraemer2020a}. Based on our assumed radial mass distribution, we used the relationship between the mass and line luminosities described in Section 3 of \citetalias{trindadefalcao2020a} to derive the radial flux distributions. Specifically, 
\begin{equation}
    L_{_{NeIX}} = \frac{M(r) F_{c}}{n_{x}\delta r \mu m_{p}}
\end{equation}
where $L_{_{NeIX}}$ is the Ne~IX luminosity calculated using Cloudy, $M(r)$ is the mass of gas in an annulus at radius $r$ derived from the [O~III] analysis, $F_{c}$ is the Ne~IX luminosity per ${\rm cm^{2}}$ calculated by Cloudy, $\delta r$ is the extraction size of the bin, $\mu$ is the mean mass per proton, and $m_{p}$ is the mass of a proton. Our results are shown in the left panel of Figure \ref{fig:NeIX_flux}.\par

The energy band-pass used to extract the Ne~IX flux in the ACIS image is 0.848 keV - 0.919 keV. Based on our models, the predicted Cloudy spatially-integrated flux for the Ne~IX triplet\footnote{Our results for the Ne~IX are given as a combination of the three lines for a better comparison to the \textit{Chandra}/ACIS image, since the individual lines cannot be spectrally resolved.} (the resonance line, 13.50\AA, the intercombination line, 13.60\AA, and the forbidden line, 13.73\AA) is $2.6^{+7.8}_{-1.9}\times 10^{-15}~{\rm erg~cm^{-2}~s^{-1}}$ (see Figure \ref{fig:NeIX_flux}). We compare our result to the observed flux for the Ne~IX derived from the \textit{Chandra}/ACIS image of Mrk 34, $\sim$ $2.6^{+2.6}_{-1.3}\times 10^{-15}~{\rm erg~cm^{-2}~s^{-1}}$, which agrees very well with the model prediction. In addition, the spatial distribution of the \textit{Chandra}/ACIS data is in rough agreement with our predicted radial flux distribution, as shown in Figure \ref{fig:NeIX_flux} (right panel). It should be noted here that our model overpredicts the flux of the central bin compared to the \textit{Chandra} data. One possibility for the discrepancy is due to the point-spread function of the ACIS image, which could result in $\sim$ 50\% of the counts registered in adjoining bonds. Another possibility is that there is absorption along the line-of-sight to the nucleus, which could result in a lower observed count rate \citep{kraemer2011a}.\par 

It is important to note that if there is interaction between the X-ray wind and the [O~III] gas it is likely that shocks are occurring. In fact, based on \textit{Chandra}/ACIS imaging, \citet{maksym2019a} determined that shocked gas is present in the NLR of the Seyfert 2 galaxy NGC 3393. However, the good agreement between the Cloudy model predictions for Ne~IX and the observed fluxes suggests that most of the X-ray emission is from photoionised gas and any shocks regions were localised, as in the Seyfert 2 NGC 1068 \citep{kraemer2000a}.\par

Our predicted integrated flux for O~VII-f, i.e., the O~VII forbidden line, 22.19\AA, is $8.30^{+24.9}_{-6.2}\times10^{-15}~{\rm erg~cm^{-2}~s^{-1}}$. The ratio of O~VII-f to NeIX r+i+f derived from our analysis is 3.2, which can be compared to those observed in other AGN. For example the ratios in NGC 4151, NGC 1068, and NGC 3516 are 2.6 \citep{armentrout2007a}, 1.7 \citep{kallman2014a} and 3.5 \citep{turner2003a}, respectively. The average of these is 2.6, which is roughly the same as our value. It is important to note that, due to the loss of sensitivity of the ACIS at low energies, we could not obtain an accurate O~VII flux.\par 

Taking into account our results for O~VII and Ne~IX for Mrk 34, and how well the latter agrees with the value derived from the \textit{Chandra}/ACIS imaging analysis, there appears to be enough X-ray emitting gas to produce the observed dynamic effects. However, there does not appear to be significantly more mass in the X-ray gas than in the [O~III] gas. Nevertheless, if the expansion of [O~III] gas is the source of the X-ray winds, one might expect that the two components could have similar masses. \par 

\subsection{Footprints of X-ray Wind}
\label{sec:footprints}

\begin{figure}
  \centering
  \includegraphics[width=0.5\textwidth, height=5.8cm]{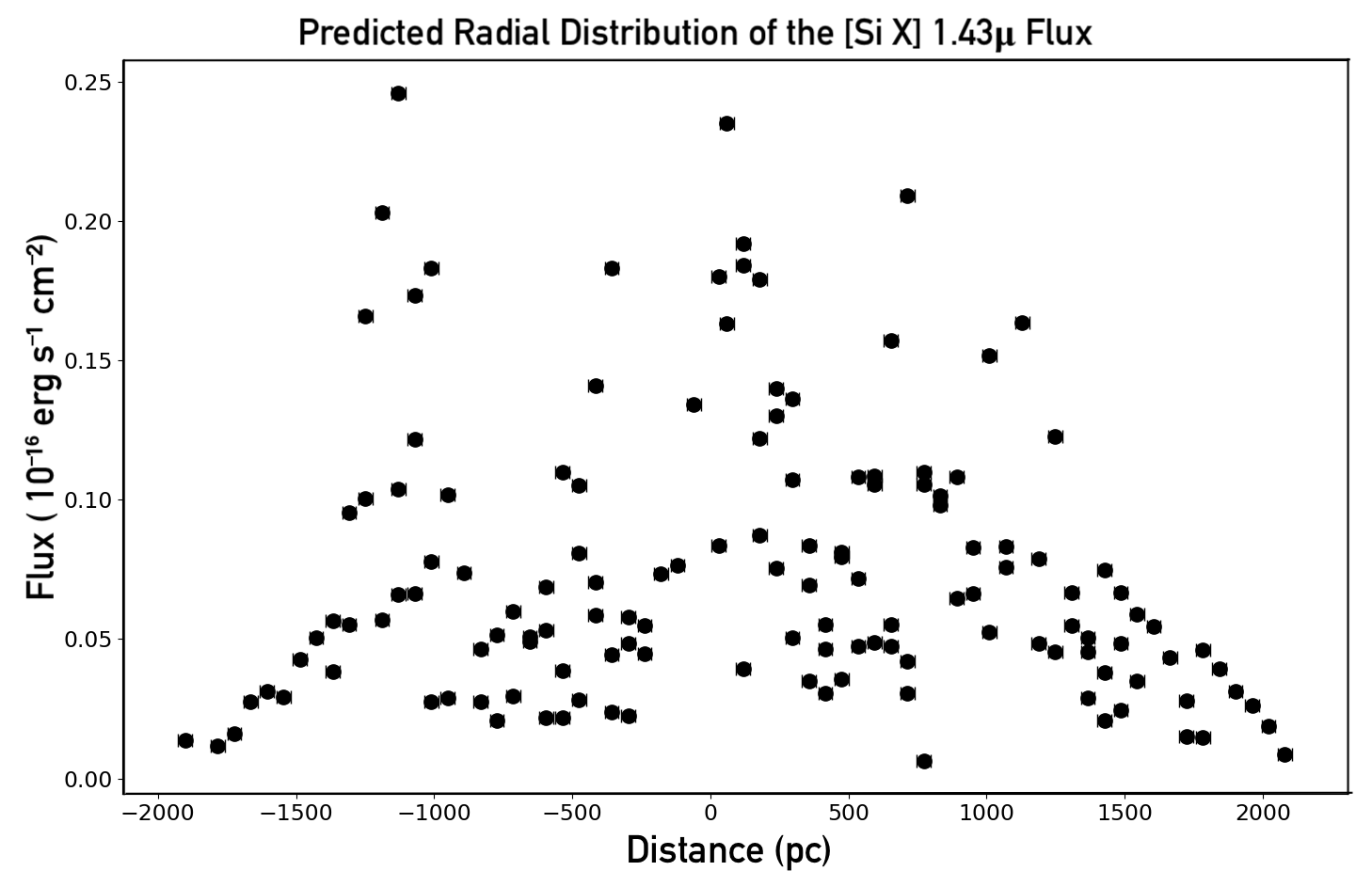}
\caption{\label{fig:SiX_flux} Radial predicted distribution of the [Si~X] flux for Mrk 34. The fluxes were obtained using the same analysis described in Paper I, section 3, and we assume that the mass of [O~III]-emitting gas and O~VII-emitting gas are roughly the same. These models are derived from the same models and mass distribution used for Ne~IX. }
\end{figure}

A model for X-ray gas, characterised by a $logU=0.0$ predicts a strong emission from optical and IR lines, which can be considered as "footprints" of the X-ray wind \citep[e.g.,][]{porquet1999a}. One of these is [Si~X] 1.43$\mu$m. Since it could be detected using ground-based near-infrared telescopes \citep[e.g.,][]{rodriguez-ardila2011a, lamperti2017a} or the \textit{James Webb Space Telescope}, it provides an opportunity to accurately constrain the kinematics of the X-ray wind. The [Si~X] fluxes and radial flux distributions were derived from the same models and mass distribution used for Ne~IX. Our results are shown in Figure \ref{fig:SiX_flux}. The spatially-integrated flux for [Si~X] 1.43$\mu$m is $1.0^{+3}_{-0.75}\times10^{-15}~{\rm erg~cm^{-2}~s^{-1}}$, which is in general agreement with the values for the [Si~X] flux found by \citet{rodriguez-ardila2011a} for their sample of nearby AGN and, therefore, it should be detectable in near-IR spectra of Mrk 34.\par 

\begin{figure}
  \centering
  \includegraphics[width=0.5\textwidth, height=5.8cm]{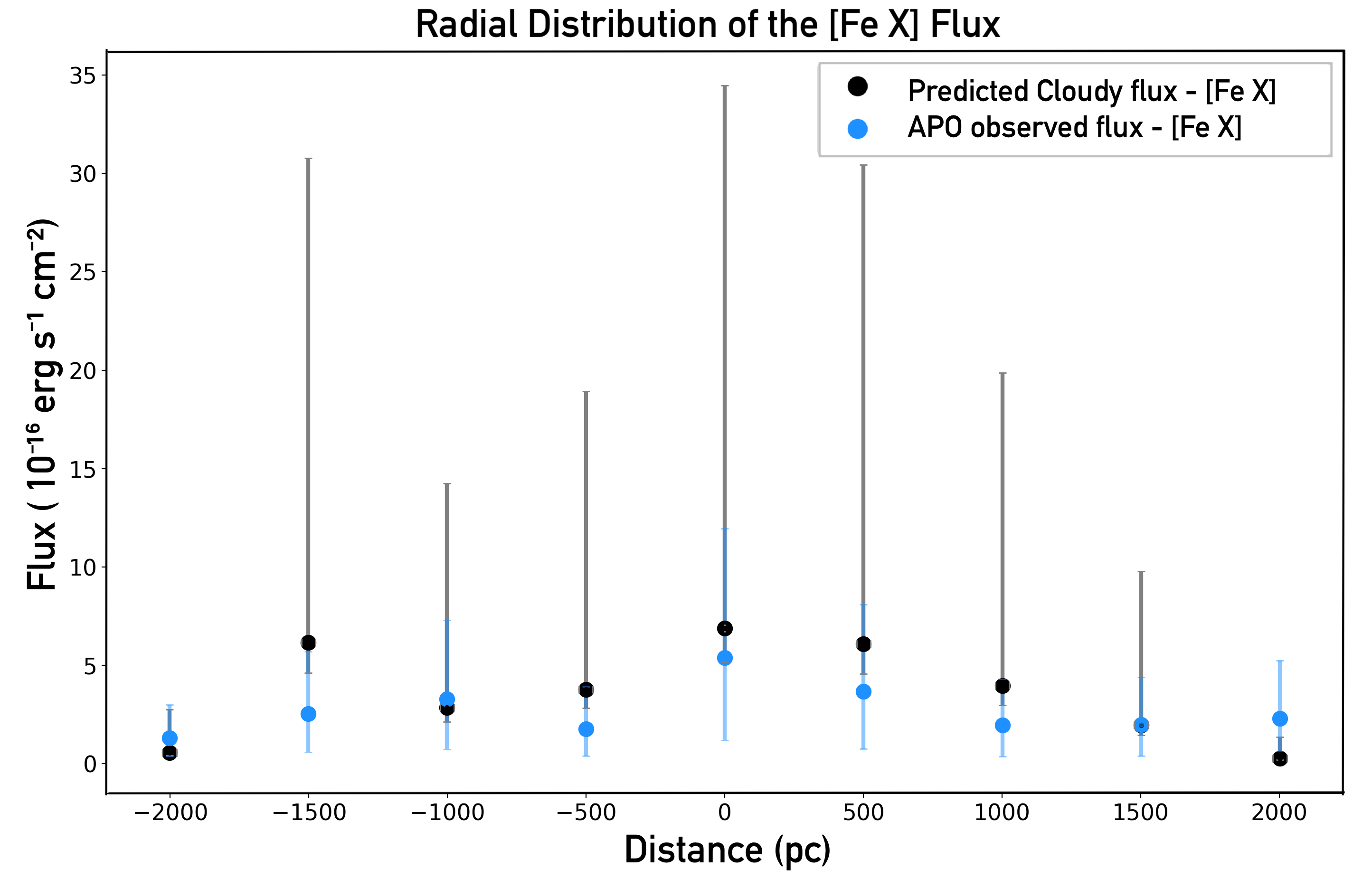}
\caption{\label{fig:FeX_flux} Comparison between the predicted Cloudy spatially-integrated flux for the [Fe~X] 6374 \AA~and the observed flux derived from the APO data for Mrk 34. The error bars in our models were derived as described in \citetalias{trindadefalcao2020a}. }
\end{figure}

 In addition, our models predict a spatially-integrated flux for the [Fe~X] 6374~\AA~in Mrk 34 of $3.2^{+9.6}_{-2.4}\times10^{-15}~{\rm erg~cm^{-2}~s^{-1}}$. Our predicted value for the [Fe~X] can be compared to the measured Apache Point Observatory (APO) data for Mrk 34 \citep{revalski2018a}. The upper limit on the integrated line flux across the entire NLR is $\sim$ $4.4\times10^{-15}~{\rm erg~s^{-1}~cm^{-2}}$ with an uncertainty of $\sim$ 30 - 50\%. Although the fit spans +/- 8", non-zero flux was only present inside +/- 4". However, the [Fe~X] line is only detected with any sort of real significance inside of $\sim$2.4". Therefore, the measured flux from the APO data, at $r$ $<$ 2.4", i.e., $\sim$ 2.3 kpc, is $3.0\times10^{-15}~{\rm erg~s^{-1}~cm^{-2}}$, which agrees very well with our predicted values. In addition, the spatial distribution of the APO data is also in good agreement with our predicted radial flux distribution, as shown in Figure \ref{fig:FeX_flux}. \par

\section{Discussion}
\label{sec:discussion}

In \citetalias{trindadefalcao2020a}, we found a range of $\dot E$/$L_{bol}$ between $3.4^{+10.2}_{-2.5}\times 10^{-8}$ and $4.9^{+14.7}_{-3.7}\times 10^{-4}$ for these QSO2s, as opposed to the $5\times 10^{-3}$ - $5\times 10^{-2}$ ratio required for efficient feedback \citep{dimatteo2005a, hopkins2010a}. In our current study we explore the possibility that the [O~III] gas we see in the NLR of some of these targets is expanding rapidly until it forms an X-ray wind, which could be powerful enough to produce efficient feedback. \par 

In Sections \ref{sec:kinetic_energy} and \ref{sec:entrainment} we present evidence of the presence of these X-ray winds in Mrk 34, which we believe are responsible for disturbing the [O~III] gas observed outside of the outflow region  \citep{fischer2018a} and for entraining [O~III] gas at $\sim$ 1.2 kpc. This process of entrainment is consistent with the outflow radial profile for Mrk 34 discussed by \citet{revalski2018a}. In Section \ref{sec:Xray_wind}, we determine the mass and mass distribution of the X-ray gas. We find that there appears to be enough X-ray gas to produce the dynamical effects we observe in Mrk 34. In order to investigate the role of X-ray winds in AGN feedback, we estimate the mass outflow rate and kinetic luminosity for the wind. \par

It is important to note here that, although our X-ray model is in good agreement with the observations, we have no independent means to measure the X-ray kinematics. However, if we assume that the X-ray emission-line gas is producing the effects  discussed in sections \ref{sec:kinetic_energy} and \ref{sec:entrainment}, we can estimate the dynamical quantities of the wind. The maximum mass outflow rate for the predicted X-ray wind is $15.2^{+45.6}_{-11.4}~{\rm M\textsubscript{\(\odot\)}}$ ${\rm yr^{-1}}$, and the maximum kinetic luminosity\footnote{The given result is for the position of maximum kinetic luminosity. However, there are several other points that fulfill the "requirement" for efficient feedback, given the mass uncertainties.} is $5.4^{+16.3}_{-4.1} \times 10^{43}~{\rm erg}~{\rm s^{-1}}$, both at $\sim$ 1.8 kpc. Comparing the kinetic luminosity to the AGN bolometric luminosity, we find a $\dot E$/$L_{bol}$ ratio of $2.1^{+6.3}_{-1.6}~\times10^{-2}$, which is consistent with the range used in models of efficient feedback \citep{dimatteo2005a, hopkins2010a}. For comparison, in \citetalias{trindadefalcao2020a} we showed that the maximum mass outflow rate for the [O~III]-emitting gas in Mrk 34 is $10.3^{+30.9}_{-7.7}~{\rm M\textsubscript{\(\odot\)}}$ ${\rm yr^{-1}}$, and the maximum kinetic luminosity of the outflows is $1.3^{+3.9}_{-0.9}~\times10^{42}{\rm erg}~{\rm s^{-1}}$ (which are in agreement with the values obtained by \citet{revalski2018a}). These results suggest that the X-ray wind may be powerful enough to generate efficient feedback and impact the host galaxy in regions where the [O~III]-emitting gas cannot.\par 

The fact that there appears to be a dynamically important X-ray wind in Mrk 34 makes us question why we do not see these same effects in lower luminosity AGNs. In NGC 4151, for example, the [O~III] observations do not show any evidence of disturbance due to X-ray winds \citep{kraemer2020a}. One possibility is that the [O~III]-emitting gas in these AGNs does not have a high enough velocity to make an efficient X-ray wind at large distances. This may occur because the $L_{bol}/L_{edd}$ ratio is too low. By looking at Figure \ref{fig:velocity_profile_3}, we can see that, in order to extend to large distances, the X-ray wind must be generated close to the AGN. However, if the AGN is not very luminous (in the case of NGC 4151, $L_{bol}/L_{edd} \sim 2$\%, as shown by \citet{kraemer2005a}), it cannot generate X-ray winds with the high velocities we see in Figure \ref{fig:velocity_profile_3}. Nevertheless, the fact that we do not see extended outflows or evidence for X-ray winds in Mrk 477 makes us question if there are other factors than $L_{bol}/L_{edd}$ ratio that determines whether these outflows occur. One possibility is that this is related to the properties of the circumnuclear gas in each AGN, which may be related to their evolutionary states. \par

As discussed by \citet{fischer2018a}, Mrk 477 possesses a very compact morphology, in which the central [O~III] emission is point-like in the narrow-band image, with fainter line emission surrounding it. One possibility is that there is more gas close to the AGN (see \citealt{fischer2018a}, their Figure 8) which results in the attenuation of a fraction of ionising radiation. Even though it is unlikely that the optical emission-line gas is responsible for attenuating much of the ionising radiation, based on its small covering factor, it is possible that this is occurring in X-ray emission-line gas. The high [O~III] FWHM close to the AGN in Mrk 477 may be evidence for interaction with an X-ray wind, as we suggest for Mrk 34.\par

In addition, as suggested by \citet{fischer2018a}, it is possible that this gas could eventually be driven away from close to the AGN, which would reduce the amount of emission-line gas at these distances. This could also drive out the high ionisation emission-line gas, i.e., X-ray gas, which would reduce the attenuation of the ionising radiation, resulting in a more extended [O III] region. This suggests that AGNs with compact emission-line morphologies, such as Mrk 477, will evolve into those with more extended emission-line regions, such as Mrk 34.\par

\section{Conclusions}
\label{sec:conclusions}
Based on the imaging and spectral analysis presented in \citetalias{trindadefalcao2020a}, we analyze the dynamics of mass outflows in the NLR of Mrk 477 and Mrk 34. Our main conclusions are as follows:

1. From our models for Mrk 477, the observed [O~III]-emitting gas was launched within 200 pc from where it is being detected. We show that there needs to be internal dust for the gas to be efficiently accelerated. Also, the outflows in this target do not extend far into its NLR. \par

2. Based on our models, in order to achieve the velocities within 500 pc of the SMBH we observe in Mrk 34, it is likely that the dusty [O~III]-emitting gas would have to start in a lower ionisation, higher density state and, then, evolve into [O~III] gas. This scenario is consistent with in-situ acceleration \citep{crenshaw2015a,fischer2017a}.\par

3. We present evidence for the presence of X-ray winds in the NLR of Mrk 34, as follows:\par 
a) We first study the possibility that these winds may be disturbing the [O~III] gas we observe outside of the outflow region in Mrk 34. We perform an analysis based on the kinetic energy density of the [O~III]-emitting gas and the X-ray winds. We conclude that, in order to have sufficient kinetic energy density, these X-ray winds would have velocities greater than the velocity dispersion in the disturbed [O~III] gas. To achieve these high velocities, the X-ray wind must originate much closer to the SMBH, in the form of lower-ionisation gas.\par 

b)We also observe high velocity [O~III]-emitting gas at $\sim$ 1.2 kpc in the NLR of Mrk 34. We believe this high velocity gas is being accelerated in-situ via entrainment by X-ray winds which originate close to the AGN, similar to the winds that cause the disturbed kinematics at greater distances. \par

4. We estimate the amount of X-ray emitting gas by comparing the integrated Ne~IX flux to that measured in the \textit{Chandra}/ACIS data. The excellent agreement indicates that our estimate of the mass of the X-ray emitting gas is generally accurate and that there appears to be enough X-ray gas to produce the dynamical effects we observe.  \par 

5. We calculate the integrated fluxes for the X-ray footprint lines, i.e., [Si~X] 1.43$\mu$m and [Fe~X] 6374~\AA. We find that our integrated and radial flux distributions for the [Fe~X] emission-line agree very well with the measured values. These results not only support our parametrization of the X-ray wind, but also demonstrate that optical and IR emission lines are "footprints" of the X-ray gas, and can be used to constrain the X-ray wind dynamics. \par

6. Our estimate of the kinetic luminosity of the X-ray wind in Mrk 34 is $2.1^{+6.3}_{-1.6}~\times10^{-2}$ of Mrk 34's bolometric luminosity, as opposed to $4.9^{+14.7}_{-3.7}$ $\times10^{-4}$ for the [O~III]-emitting gas (see \citetalias{trindadefalcao2020a}). These estimates are based on the assumption that the X-ray emitting gas is the same material that produces the observed dynamical effects. This result indicates that the X-ray wind, being more powerful by a factor of 50, may be an efficient feedback mechanism, from the criteria of \citet{dimatteo2005a} and \citet{hopkins2010a}. \par 

This shows that the outflow X-ray wind has the potential for affecting the host galaxy and to play a greater role in AGN feedback than the optical emission-line gas (\citetalias{trindadefalcao2020a}). This would then occur in targets that are radiating near their Eddington limit \citep[e.g.,][]{fischer2019a}. Our study shows that the interaction with X-ray winds could be the source of the dynamic phenomena we observe in luminous AGN, such as Mrk 34. \par

Several issues have not been fully explored in this study, such as the variations in the internal dust/gas ratios (see section \ref{sec:force_multiplier}) and the physical structure of the clouds \citep[e.g.,][]{chelouche2001a}. However, the main open issues concern cloud confinement and stability. We present possible explanations to the impasse of instability, but a resolution must be left to detailed hydrodynamic models \citep[e.g.,][]{zhang2017a,gronke2019a}. Nevertheless, the results of our dynamical analysis, such as radial velocity profiles, acceleration timescales, and the role of the X-ray, can provide valuable new constraints for more sophisticated modeling.\par

\section*{Acknowledgements}

The authors thank the anonymous referee for helpful comments that improved the clarity of this paper. Support for this work was provided by NASA through grant number HST-GO-13728.001-A from the Space Telescope Science Institute, which is operated by AURA, Inc., under NASA contract NAS 5-26555. Basic research at the Naval Research Laboratory is funded by 6.1 base funding. T.C.F. was supported by an appointment to the NASA Postdoctoral Program at the NASA Goddard Space Flight Center, administered by the Universities Space Research Association under contract with NASA. T.S.-B. acknowledges support from the Brazilian institutions CNPq (Conselho Nacional de Desenvolvimento Científico e Tecnológico) and FAPERGS (Fundação de Amparo à Pesquisa do Estado do Rio Grande do Sul). L.C.H. was supported by the National Key R\&D Program of China (2016YFA0400702) and the National Science Foundation of China (11721303, 11991052). M.V. gratefully acknowledges financial support from the Danish Council for Independent Research via grant no. DFF 4002-00275 and 8021-00130.\par 
This research has made use of the NASA/IPAC Extragalactic Database (NED), which is operated by the Jet Propulsion Laboratory, California Institute of Technology,
under contract with the National Aeronautics and Space Administration. This paper used the photoionisation code Cloudy, which can be obtained from http://www.nublado.org. We thank Gary Ferland and associates, for the maintenance and development of Cloudy. In addition, the authors thank L. M. Laramee for help with the manuscript.

\section*{Data Availability}

Based on observations made with the NASA/ESA Hubble Space Telescope, and available from the Hubble Legacy Archive, which is a collaboration between the Space Telescope Science Institute (STScI/NASA), the Space Telescope European Coordinating Facility (ST-ECF/ESAC/ESA) and the Canadian Astronomy Data Centre (CADC/NRC/CSA).



\bibliographystyle{mnras}
\bibliography{anna_bibliography} 








\bsp	
\label{lastpage}
\end{document}